\documentclass{article}

\usepackage{ifthen}
\newboolean{@EDMGRVersion}
\newboolean{@PlainVersion}

\setboolean{@EDMGRVersion}{false}  
\setboolean{@PlainVersion}{true} 

\ifthenelse{\boolean{@PlainVersion}}{%
  
  \newcommand{\tabcapfont}{}
  \newcommand{\inlinecite}[1]{\cite{#1}}
}{
  \usepackage{klucite}
}
\usepackage{amsmath}
\usepackage{graphicx}
\usepackage{url}
\usepackage{xspace}
\usepackage{array,supertabular}
\usepackage{chemarr}

\newcommand{\CSH}{\mbox{C-S-H}\xspace}
\newcommand{\CtwoS}{{\mbox{C$_2$S}}}
\newcommand{\CthreeS}{{\mbox{C$_3$S}}}
\newcommand{\CthreeA}{{\mbox{C$_3$A}}}
\newcommand{\CaO}{{\mbox{CaO}}}
\newcommand{\SiO}{{\mbox{SiO$_2$}}}
\newcommand{\AlO}{{\mbox{Al$_2$O$_3$}}}

\newcommand{\HO}{{\mbox{H$_2$O}}}
\newcommand{\CH}{\mbox{CH}\xspace}
\newcommand{\OH}{{\mbox{OH$^{-}$}}}
\newcommand{\Ca}{{\mbox{Ca$^{2+}$}}}
\newcommand{\HSO}{{\mbox{H$_2$SiO$_{4}^{2-}$}}}
\newcommand{\cC}{{\mbox{C}}}
\newcommand{\cS}{{\mbox{S}}}
\newcommand{\cH}{{\mbox{H}}}
\newcommand{\leavethisout}[1]{}

\newcommand{\en}[1]{(\ref{eq:#1})}
\newcommand{\water}{w} 
\newcommand{\gel}{g}   
\newcommand{\sat}{\theta}
\newcommand{\satmin}{\sat_{\text{\emph{min}}}}
\newcommand{\satmax}{\sat_{\text{\emph{max}}}}
\newcommand{\satrx}{\sat_r}
\newcommand{\sattol}{0.0005}
\newcommand{\satfac}{(\sat-\satrx)^+}
\newcommand{\Rwc}{R_{w/c}}
\newcommand{\Rac}{R_{a/c}}
\newcommand{\alite}{\alpha}
\newcommand{\belite}{\beta}
\newcommand{\aq}{q}
\newcommand{\Calite}{C_\alite}
\newcommand{\Cbelite}{C_\belite}
\newcommand{\CCSHaq}{C_\aq}
\newcommand{\CCSHgel}{C_\gel}
\newcommand{\rhoCSHgel}{\rho_\gel}
\newcommand{\rhoagg}{\rho_{\text{\emph{agg}}}}
\newcommand{\rhocem}{\rho_{\text{\emph{cem}}}}
\newcommand{\rhowater}{\rho_\water}
\newcommand{\Mwater}{m_\water}
\newcommand{\mday}{\mbox{\emph{day}}}

\newcommand{\constit}{j}
\newcommand{\vol}{\varepsilon}
\newcommand{\porosity}{\vol}
\newcommand{\porref}{\porosity^o}

\newcommand{\kads}{k_{\text{\emph{prec}}}}
\newcommand{\kdes}{k_{\text{\emph{diss}}}}
\newcommand{\units}[1]{\mbox{$[#1]$}}
\newcommand{\MCSH}{m_{\text{\emph{csh}}}}
\newcommand{\rCSH}{r_{\text{\emph{csh}}}} 
\newcommand{\Xmax}{L}
\newcommand{\dx}{h} 
\newcommand{\stoich}{\nu}
\newcommand{\Comix}{C_{\text{\emph{mix}}}}
\newcommand{\CSHformula}{\mbox{C$_3$S$_2$H$_3$}}
\newcommand{\etal}{{et~al.}}
\newcommand{\myparagraph}[1]{\par\vspace*{0.2cm}\emph{#1.}\ \ }

\newcommand{\myepsfile}[3]{%
  \ifthenelse{\boolean{@EDMGRVersion}}{%
    \includegraphics[width=#3]{#2} 
  }{
    \ifthenelse{\boolean{@PlainVersion}}{%
      \includegraphics[width=#3]{#2} 
    }{
      \includegraphics[width=#3]{#1} 
    }
  }
}

\newcommand{\mymaclaptop}{Macintosh PowerBook with a 1.67 GHz PowerPC G4 
  processor}

\newenvironment{mynote}{
  \begin{quote}\begin{description}\item[{\bfseries\itshape Note:}]
      \itshape
}{\end{description}\end{quote}}

\newcommand{\PreserveBackslash}[1]{\let\temp=\\#1\let\\=\temp}
\let\PBS=\PreserveBackslash
\newcommand{\rrcol}{\PBS\raggedright\hspace{0pt}}

\begin{document}

\newcommand{\mytitle}{A model for reactive porous transport during
  re-wetting of hardened concrete}

\newcommand{\myabstract}{A mathematical model is developed that captures
  the transport of liquid water in hardened concrete, as well as the
  chemical reactions that occur between the imbibed water and the
  residual calcium silicate compounds residing in the porous concrete
  matrix.  The main hypothesis in this model is that the reaction
  product -- calcium silicate hydrate gel -- clogs the pores within the
  concrete thereby hindering water transport.  Numerical simulations are
  employed to determine the sensitivity of the model solution to changes
  in various physical parameters, and compare to experimental results
  available in the literature.}

\ifthenelse{\boolean{@PlainVersion}}{%
  
  \title{\mytitle}

  \author{Michael Chapwanya\footnotemark[2]
      \and John M.\ Stockie\footnotemark[2]\;$^{,}$\footnotemark[4]\  
      \and Wentao Liu\footnotemark[3]}

  \footnotetext[2]{Department of Mathematics, Simon Fraser University,
    Burnaby, BC, Canada.} 
  \footnotetext[3]{Department of Applied Mathematics, University of
    Waterloo, Waterloo, ON, Canada.} 
  \footnotetext[4]{Corresponding author (stockie@math.sfu.ca).}

  \maketitle
  
  \section*{Abstract}
  \myabstract

}{%

\begin{opening}
  
  \title{\mytitle}

  \author{Michael \surname{Chapwanya}} 
  \author{John M. \surname{Stockie}\thanks{Corresponding author
      \email{stockie@math.sfu.ca}.}}
  \institute{Department of Mathematics, Simon Fraser University, Burnaby,
    BC, Canada}
  
  \author{Wentao \surname{Liu}} 
  \institute{Department of Applied Mathematics, University of Waterloo,
    Waterloo, ON, Canada} 

  \runningtitle{A model for re-wetting of hardened concrete}

  \runningauthor{M. Chapwanya, J. M. Stockie and W. Liu}

  \begin{ao}
    Department of Mathematics\\
    Simon Fraser University\\
    8888 University Drive\\
    Burnaby, BC, V5A 1S6\\
    Canada\\
    Tel: +1 778 782 3553\\
    Fax: +1 778 782 4947
  \end{ao}

  \begin{abstract}
    \myabstract
  \end{abstract}



\end{opening}
}

\section{Introduction}
\label{sec:intro}

Concrete is a ubiquitous construction material that derives its utility
from a combination of strength, versatility and relatively low cost.  In
fact, concrete is the second most consumed material on the planet next
to water~\cite{wbcsd-report-2002}.  The primary ingredients that go
into the making of concrete are water, Portland cement (a fine powder
consisting primarily of calcium silicate compounds), and solid
aggregates such as sand and gravel.  When mixed together, these
ingredients undergo a complex physico-chemical transformation which can
be divided into a number of discrete steps: an initial hydration stage
that occurs over a period of hours or days; a drying/curing period that
can require months or even years to complete; and additional reactions
arising from carbonation and various degradation processes that
typically also occur over very long time periods.

Mathematical modelling of transport and reaction in concrete has been
the subject of a large number of papers in the scientific and
engineering literature.  The earliest study that we are aware of which
treats water transport in concrete as a nonlinear diffusion process is
that of Ba\v{z}ant and Najjar~\cite{bazant-najjar-1971}.  Later work
considered the additional effect of transport and reaction of chemical
species in the context of initial cement
hydration~\cite{billingham-etal-2005,preece-billingham-king-2001,bary-sellier-2004}
or concrete
carbonation~\cite{papadakis-vayenas-fardis-1989,meier-etal-2007,ferretti-bazant-2006}.
We remark that most of these models assume a constant porosity even
though experimental evidence overwhelmingly suggests that the pore
structure varies significantly over time owing to reactant consumption,
crys\-tallization, and swelling of products throughout the various
stages of concrete
hydration~\cite{papadakis-vayenas-fardis-1989,hall-etal-1995,taylor-etal-1999}.
In fact, the only models we are aware of that allow for a variable
porosity are in the context of carbonation, where
Meier \etal~\cite{meier-etal-2007} specify the porosity as a given
decaying exponential function of time, while Bary and
Sellier~\cite{bary-sellier-2004} allow the porosity to depend on the
solution via changes in the pore volume from solidified reaction
products.

We focus here on a later stage in the life of concrete, namely the
process of re-wetting or ``secondary hydration'' in which hardened and
cured concrete experiences imbibition of water, due to periodic rainfall
for example.  The proportion of reactive silicates in the cement that
are consumed during the initial hydration reactions (called the
\emph{degree of hydration}) is typically on the order of
50\%~\cite{bentur-2002}; consequently, there are significant levels of
residual reactants remaining in hardened concrete and so the effect of
secondary reactions occurring during re-wetting cannot be ignored.  This
study is motivated by the experimental work of
Barrita \etal~\cite{barrita-2002,barrita-etal-2004} and
Taylor \etal~\cite{taylor-etal-1999} who placed dry concrete samples in
a liquid bath and carefully observed the progress of the subsequent
wetting front.  They found that when a non-reactive liquid such as
isopropanol was used, the front speed was proportional to the square
root of time as predicted by nonlinear diffusion analysis.  When water
was used instead, the wetting front moved more slowly than the theory
predicted and in some cases stalled completely -- an effect that is
usually referred to as \emph{anomalous diffusion}.
Hall~\cite{hall-2007} has suggested that this effect is due to
physico-chemical interactions between the wetting fluid and the porous
solid.  It is natural to hypothesize therefore that the reduction in
wetting front speed arises from residual calcium silicates in the porous
matrix reacting with water to form calcium silicate hydrate or \CSH,
which precipitates in the form of a gel that clogs the pores in the
concrete; this hypothesis is supported by the results
in~\cite{hall-etal-1995}. 

Observations of anomalous diffusion have been reported
in~\cite{kuntz-lavallee-2001} where the authors proposed instead that
deviations in wetting front speed can be modeled using a modified
(non-Darcian) porous transport equation.  This approach provides a
reasonable match with experiments and gives rise to a new and
potentially interesting class of nonlinear diffusion equations and
scaling laws; however, there is no direct support for this model in
terms of a physical mechanism for concrete hydration.  In a related
study~\cite{lockington-parlange-2003}, another model is proposed
that includes an explicit time-dependence in the water
diffusion coefficient.  They showed that by assuming the cumulative
deposition follows a power law in time, they could reproduce similar
clogging results; unfortunately, it is not at all clear how one would
obtain the power law coefficients in a given wetting scenario.

Some authors have addressed clogging phenomena in the context of
concrete carbonation, such as Saetta \etal~\cite{saetta-etal-1995} who
incorporated a functional dependence on the carbonate concentration into
the transport coefficients of their model.  Meier
\etal~\cite{meier-etal-2007} also employed an empirical approach, but
instead they assumed the porosity decays as a given exponential function
of time, which has the disadvantage that there is no direct coupling
between water transport and the precipitated reaction products that are
causing the actual clogging.  Related work on self-desiccation (or
internal drying) during initial hydration and its connection with
autogenous shrinkage have been studied using pore-level microstructure
simulations~\cite{bentz-1999}.

In this paper, we develop a model that aims to test the hypothesis that
incorporating the chemistry of residual cement constituents and the
effect of the resulting \CSH gel formation on pore structure can explain
the apparent clogging effects observed in concrete re-wetting
experiments.  We begin in Section~\ref{sec:chemistry} with a brief
overview of cement chemistry and the physico-chemical changes that occur
in cement during hydration.  We develop the mathematical model in
Section~\ref{sec:model} using a macroscopic approach that is motivated
by the clogging models developed in the bioremediation literature (see
for example \cite{clement-hooker-skeen-1996}) wherein the
accumulation of biomass -- analogous to cement hydration products -- is
responsible for the reduction in porosity.  Numerical simulations of the
resulting system of nonlinear partial differential equations are
performed in Section~\ref{sec:numerics} and the results are compared
with experiments.  We show that our model captures observed clogging
behaviour both qualitatively and quantitatively with a minimum of
parameter fitting, and we explain in Section~\ref{sec:conclusion} how
these results might be generalized in future to handle a range of other
related phenomena in concrete transport.

\section{Overview of cement chemistry}
\label{sec:chemistry}

While this paper is not concerned directly with the primary hydration of
cement, the same hydration reactions occur during the re-wetting phase
when residual unhydrated silicates remaining in the hardened concrete
matrix are exposed to water.  We will therefore begin by presenting some
background information on the process of cement hydration that is drawn
largely from~\cite{lea-1970,bentz-1995}.  Portland cement is
the key binding agent in concrete and has as its major constituents
tricalcium silicate ($3\CaO\cdot\SiO$, commonly referred to as
alite) and dicalcium silicate ($2\CaO\cdot\SiO$, known as
belite) which make up approximately 50\%\ and 25\%\ respectively of
dry cement by mass.  The remaining 25\%\ consists primarily of
tricalcium aluminate, tetracalcium aluminoferrite, and gypsum, with
smaller amounts of certain other admixtures whose purpose is to
influence such properties as strength, flexibility, setting time, etc.
In this paper we will concentrate solely on the two primary
constituents, alite and belite.

The cement powder is mixed with water and aggregates (sand, gravel and
crushed stone) to make a workable paste that can then be easily poured
or molded and left to harden.  During the initial hydration stage, the
silicates dissolve in and react with the water to form calcium hydroxide
or \mbox{Ca(OH)$_2$}, and calcium silicate hydrate or \CSH; the
latter notation does not denote a specific chemical compound but rather
represents a whole family of hydrates having Ca/Si ratios that range
between 0.6 and 2.0.  A significant amount of heat is released during
the conversion of alite and belite into \CSH since the hydration
reactions are exothermic.  Calcium hydroxide and \CSH precipitate out of
solution in crystalline form, and these solid precipitates then act as
nucleation sites that further enhance formation of \CSH.  It is the
crystalline or gel form of \CSH that is ultimately responsible for
the strength of concrete.


The hydration process can proceed via several possible reactions, but we
restrict ourselves to a particular reaction sequence that is employed in
both~\cite{preece-billingham-king-2001}
and~\cite{tzschichholz-zanni-2001}.  The mechanism for alite hydration
begins with a dissolution phase 
\begin{gather*}
  3\CaO\cdot \SiO + 3\HO \stackrel{r_1}{\longrightarrow} 3\Ca +
  4\OH + \HSO,
\end{gather*}
followed by a reaction in solution to form aqueous \CSH
\begin{gather*}
  \HSO + \frac{3}{2}\Ca + \OH
  \stackrel{r_2}{\longrightarrow} \CSH~(aq),
\end{gather*}
and precipitation of calcium hydroxide according to
\begin{gather*}
  \Ca + 2\OH \stackrel{r_3}{\longrightarrow} \mbox{Ca}(\mbox{OH})_2
\end{gather*}
In each chemical formula we have indicated the rate of the reaction by
$r_i$ \units{\mday^{-1}} for $i=1,2,3$.

For the remainder of this paper, we adopt the cement chemistry
convention in which the following abbreviations are used: $\cC=\CaO$,
$\cS=\SiO$ and $\cH=\HO$.  Then the overall reaction, leaving out
intermediate ionic species, can be written in terms of the single
formula
\begin{gather}
  2\CthreeS + 6\cH \stackrel{ r_{\alite} }{\longrightarrow} \CSH~(aq)
  + 3\CH,
  \label{eq:rx-alite}
\end{gather}
where $r_{\alite}$ represents an overall rate constant for alite.
Motivated by the models developed
in~\cite{papadakis-vayenas-fardis-1989,meier-etal-2007,bentz-1995}, we
consider only the simplified kinetics represented by \en{rx-alite}.  We
also take the chemical form of \CSH to be that of $\CSHformula$, that
because of the amorphous nature of \CSH can only be true in some
averaged sense.  A similar formula holds for belite
\begin{gather}
  2\CtwoS + 4\cH \stackrel{r_{\belite}}{\longrightarrow} \CSH~(aq) +
  \CH, \label{eq:rx-belite}
\end{gather}
where we note that $r_\alite\gg
r_\belite$~\cite{billingham-etal-2005,papadakis-vayenas-fardis-1989,bentur-2002}.
Alite is also mainly responsible for the early stage strength of the
concrete (through approximately the first seven days) while belite
contributes to the later strength.



Following~\cite{tzschichholz-zanni-2001}, we include a precipitation (or
deposition) step in which the aqueous \CSH product precipitates out of
solution to bind with the porous matrix:
\begin{gather}
    \CSH~(aq)~~
    \xrightleftharpoons[\kdes]{\kads}
    ~~\CSH~(gel) ,
    \label{eq:csh-ads}
\end{gather}
where the rate of precipitation is denoted by $\kads$
\units{\mday^{-1}}.  We allow for a dissolution process with rate
constant $\kdes$, although in most of our later simulations we restrict
ourselves to $\kdes=0$ so as to be consistent with other models
that disregard the effect of \CSH dissolution.

The hydration chemistry of other cement constituents such as aluminates,
ferrites, etc. are not considered here because they do not contribute
appreciably to the porous structure of the
concrete~\cite{billingham-etal-2005,bentz-2006}.  Instead, we focus on
the effect of \CSH gel on decreasing porosity and hindering moisture
transport within the porous concrete matrix.

\section{Mathematical model}
\label{sec:model}

\begin{figure}[tbhp]
  \tabcapfont
  \setlength{\unitlength}{1.0cm}
  \begin{picture}(5,6)
    \tabcapfont
    \put(0,3.0){\begin{tabular}{m{0.45\textwidth}cm{0.45\textwidth}}
        \myepsfile{geometry}{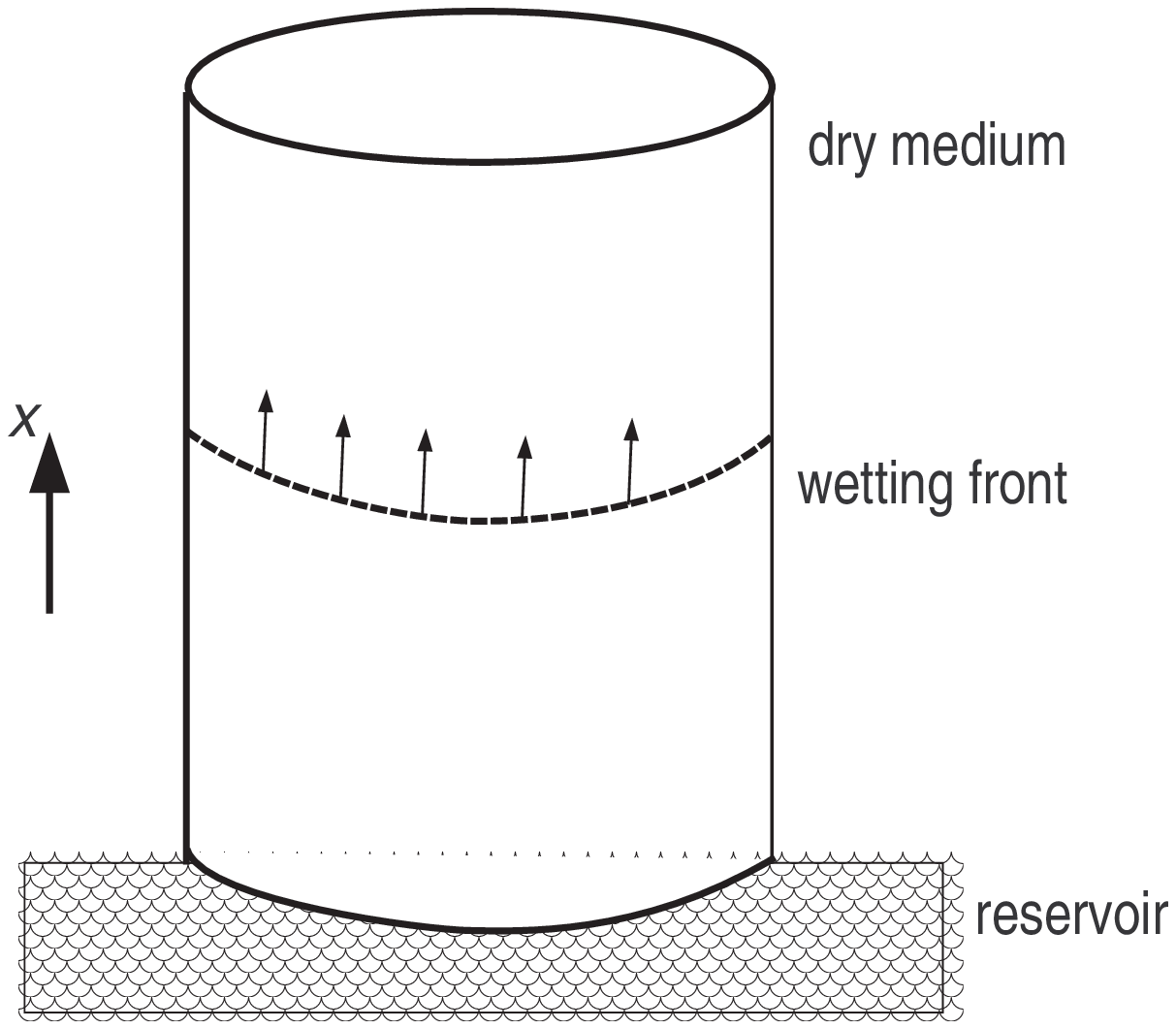}{0.45\textwidth} 
        & & 
        \myepsfile{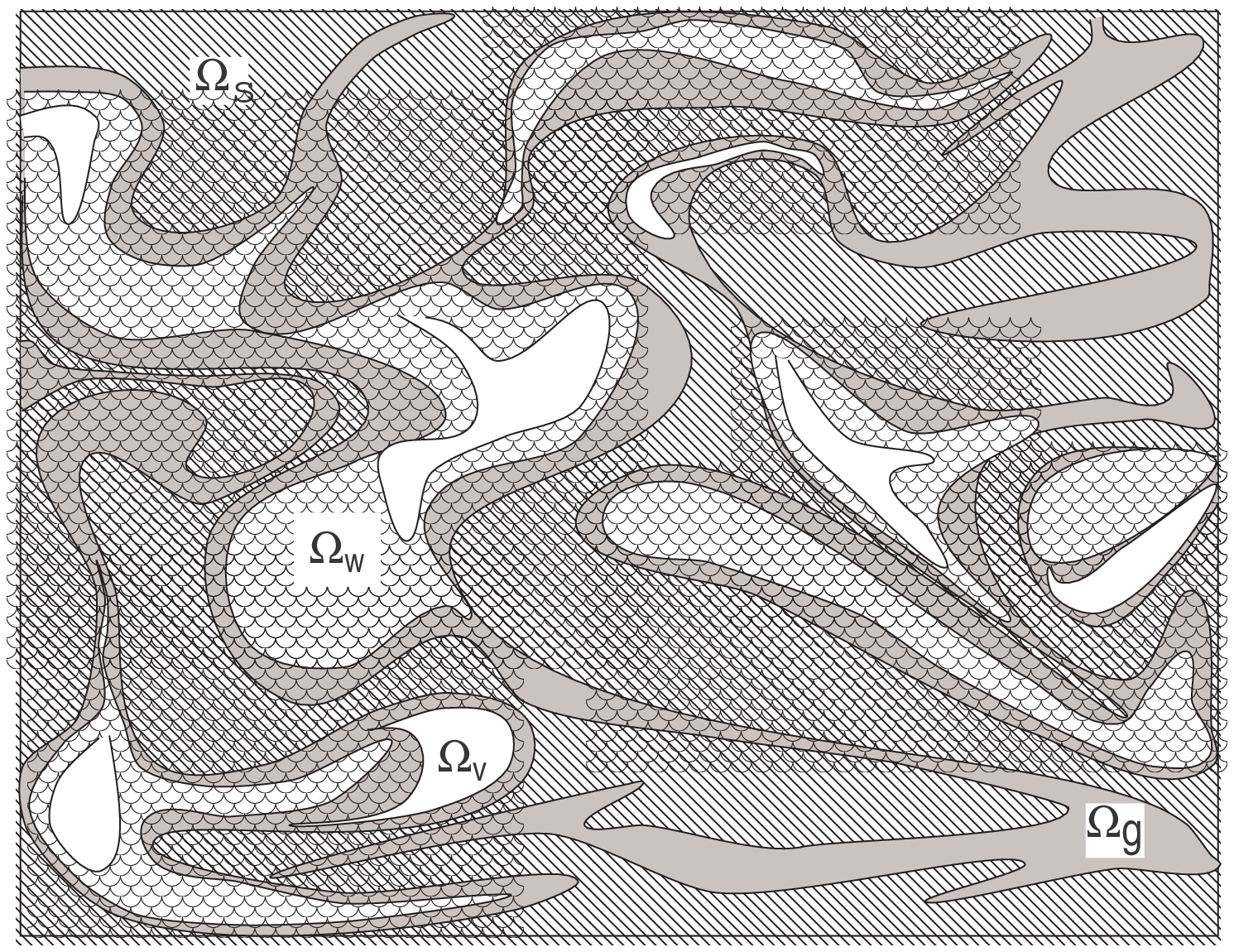}{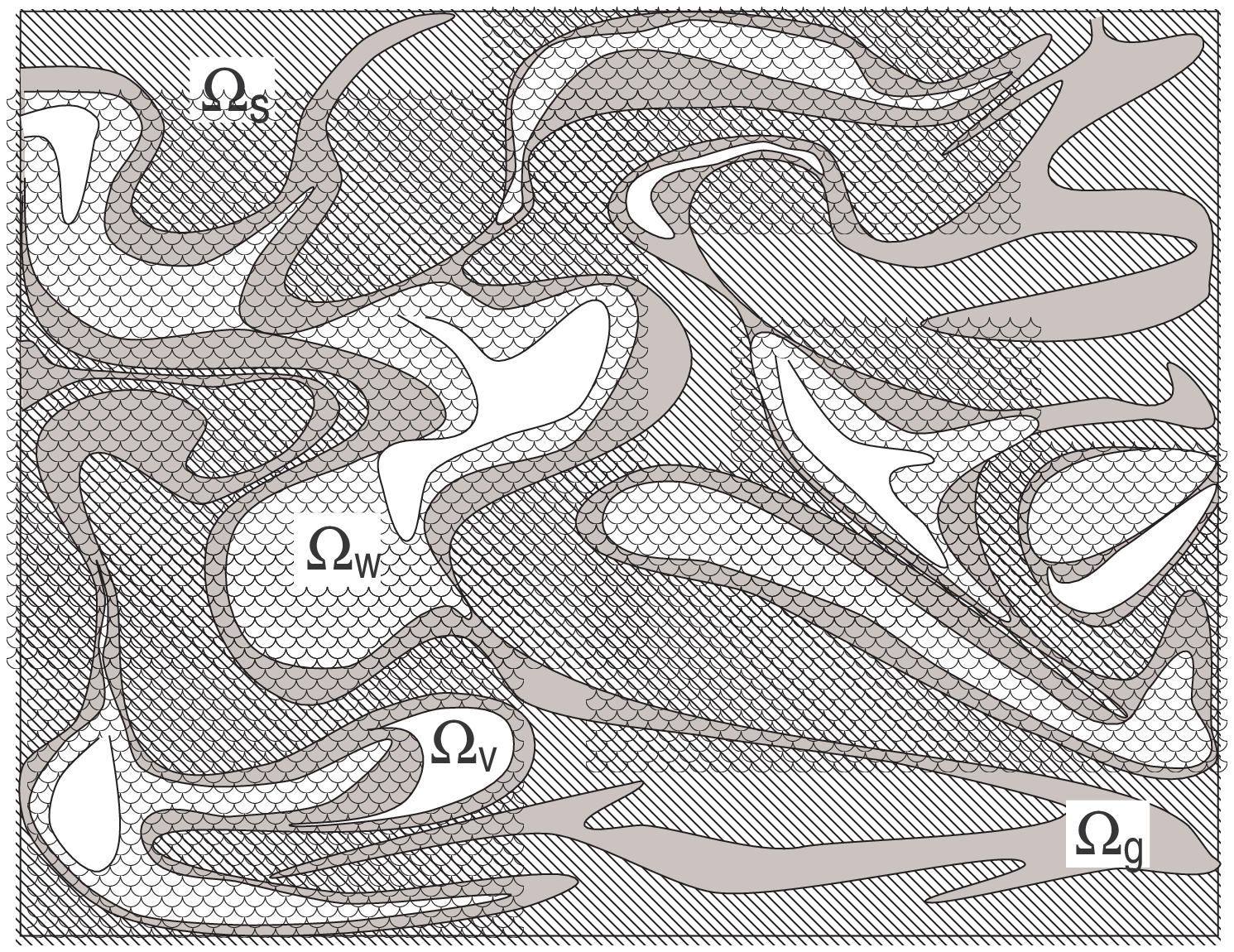}{0.45\textwidth} \\
        a.~~Cylindrical concrete sample, depicting the location of a typical
        wetting front. &&
        b.~~A zoomed-in view at the pore scale, showing a representative
        elementary volume $\Omega$.
      \end{tabular}}
    \put(3.44,3.1){\vector(1,0){3.13}}
    \put(2.8,2.8){\framebox(0.6,0.6)}
  \end{picture}
  \caption{Geometry of the 1D moisture transport problem.}
  \label{fig:geometry}
\end{figure}

We begin by providing a list of primary simplifying assumptions that
will permit us to reduce the complexity of the governing equations while
at the same time retaining the essential aspects of the underlying
physical and chemical processes:
\begin{enumerate}
\item The length scales under consideration are large enough that the
  solid concrete matrix can be treated as a continuum.  Consequently,
  volume fractions and constituent concentrations can be expressed as
  continuous functions of space and time, and the liquid transport 
  is assumed to obey Darcy's law.
\item The concrete sample is long and thin so that transport can be
  assumed to be one-dimensional.  This is consistent with many
  experiments involving concrete or other building
  materials~\cite{barrita-etal-2004} in which the sample under study
  takes the shape of a long cylinder as pictured in
  Fig.~\ref{fig:geometry}a.
\item Temperature variations and heat of reaction can be ignored.  This
  is a reasonable assumption in re-wetting of hardened concrete for
  which the quantities of silicate reactants are much smaller than
  during the initial hydration stage~\cite{hansen-1986}.
\item Water transport is dominated by capillary action and so
  gravitational effects can be ignored. This assumption is justified by
  the very small pore dimensions that lead to a small Bond number for
  concrete~\cite{lockington-parlange-2003}.
\item We neglect the dynamics of individual ionic species, which is
  consistent with the work of Bentz~\cite{bentz-1995} and others.
  Nonetheless, we do consider separate aqueous and solid phases of \CSH
  and include a simple dynamic mechanism for precipitation and
  dissolution which is shown in~\cite{tzschichholz-zanni-2001} to be an
  important effect.  This choice is motivated by the recognized
  complexity of the \CSH precipitation/crystallization
  process~\cite{grutzeck-1999}, that is largely ignored by other models
  of hydration.
\item The effects of chemical shrinkage and subsequent self-desiccation
  can be significant during initial cement hydration, particularly for
  high performance concrete~\cite{persson-1997}.  However, we neglect
  both effects since the samples under consideration are normal strength
  concrete, and residual silicate concentrations are much smaller during
  re-wetting.
\item Reaction kinetics take a simple form in which the rate has a
  power-law dependence on reactant concentration -- a common assumption
  employed in other
  models~\cite{papadakis-vayenas-fardis-1989,meier-etal-2007}.
\end{enumerate}

\subsection{Definition of volume fractions and concentrations}

Consider an elementary volume $\Omega$ \units{cm^3} pictured in
Fig.~\ref{fig:geometry}b which is divided into sub-volumes occupied by
the various components of the porous matrix, namely the non-gel solids
with volume $\Omega_s$, the precipitated \CSH gel $\Omega_\gel$, liquid
water $\Omega_\water$, and gas/vapour component $\Omega_v$.  The pore
volume available for transport is denoted by
$\Omega_p=\Omega_\water+\Omega_v$ and so the total volume can be written
as
\begin{gather*}
  \Omega=\Omega_s+\Omega_\gel+\Omega_p
  =\Omega_s+\Omega_\gel+\Omega_\water+\Omega_v.
\end{gather*}


We next define the various volume fractions beginning with the pore
volume fraction $\porosity=\Omega_p/\Omega$ \units{cm^3/cm^3}, which is
also known as porosity.  The initial porosity in the absence
of \CSH is denoted by the constant $\porref =
\porosity|_{t=0}=(\Omega-\Omega_s)/\Omega$.  The gel volume fraction is
$\vol_\gel=\Omega_\gel/\Omega$ and volumetric water content is
$\sat=\Omega_\water/\Omega$.  In practice, $\sat$ must satisfy
$0\leq\satmin\leq\sat\leq\satmax\leq \porref$, where $\satmin$ is the immobile
or residual water content and $\satmax$ is the maximum or fully
saturated value (representing the point beyond which water can no longer
penetrate the smallest pores).  In the context of cement hydration,
there are three forms of water present: chemically bound, physically
bound, and capillary water.  The quantity $\satmin$ corresponds to both
physically bound (absorbed) and capillary water, which together
represent the ``evaporable water'' that can be removed only by forced
drying.  All volume fractions will in general be functions of both
position and time owing to variations in the gel, liquid, and gas
concentrations.

We next define the concentrations of the various constituents (in units
of \units{g/cm^3}), which are measured relative to the total mass of
concrete following Ref.~\cite{papadakis-vayenas-fardis-1989}:
\begin{center}
  \begin{tabular}{cp{0.75\textwidth}}
    $\Calite(x,t)$  -- & concentration of $\CthreeS$ in concrete, \\
    $\Cbelite(x,t)$ -- & concentration of $\CtwoS$ in concrete,\\
    $\CCSHaq(x,t)$  -- & concentration of \CSH in liquid,\\
    $\CCSHgel(x,t)$ -- & concentration of solid \CSH gel
                         $=\rhoCSHgel\Omega_\gel/\Omega=\rhoCSHgel\vol_\gel$.
  \end{tabular}
\end{center}
All solution components are taken to be functions of time $t$ and 
axial distance $x$, where $t\geq 0$ and $0\leq x\leq \Xmax$.  The
gel-modified porosity $\porosity(x,t)$ is related to \CSH gel
concentration via
\begin{gather}
  \porosity = \frac{\Omega-\Omega_s-\Omega_\gel}{\Omega} =
  \frac{\Omega-\Omega_s}{\Omega} - \frac{ \Omega_\gel}{ \Omega} = \porref
  -\frac{\CCSHgel}{\rhoCSHgel},
  \label{eq:porosity}
\end{gather}
where $\rhoCSHgel$ is the density of \CSH in gel form \units{g/cm^3}.

\subsection{Derivation of the governing equations}

We next derive the differential equations governing the water content
$\sat$ and each of the constituent concentrations $\Calite$, $\Cbelite$,
$\CCSHaq$ and $\CCSHgel$.  Conservation of liquid in the pores requires
that
\begin{gather}
  \frac{\partial\sat}{\partial t} =
  -\frac{\partial u}{\partial x} - \stoich \satfac \,
  \frac{\Mwater\,\rCSH}{\rhowater\,\MCSH},
  \label{eq:sat}
\end{gather}
where $u$ is the water velocity \units{cm/\mday} and $\rhowater$ is its
density \units{g/cm^3}.  The reaction term is scaled by a factor
$\satfac=\max(\sat-\satrx,0)$ which reflects the assumption that
reactions proceed only when water content is above some minimum value
$\satrx$.  A similar approach was used in studies of concrete
carbonation~\cite{meier-etal-2007,saetta-etal-1995} wherein the value of
$\satrx$ is obtained experimentally; in the absence of experimental data
for hydration, we simply take $\satrx=\satmin$.  The expression, $\rCSH$
representing the rate of generation of \CSH \units{g/cm^3\,\mday} must
be scaled here by the ratio of the molar masses of water and \CSH,
$\Mwater/\MCSH$.  We also multiply the reaction term by a stoichiometric
coefficient $\stoich$, which is taken equal to $5$ so as to balance the
rate of generation of water with the averaged rate coefficient for \CSH
coming from Eqs.~\en{rx-alite} and \en{rx-belite}.  This and other
reaction terms are specified later in Section~\ref{sec:rates}.

We assume the liquid velocity can be expressed as
\begin{gather}
  u = - D(\sat,\porosity) \, \frac{\partial \sat}{\partial x},
  \label{eq:darcy}
\end{gather}
which follows from a simple application of Darcy's law\footnote{Darcy's
  law states that the velocity $u=K\, \partial h/\partial x$, where $h$
  is the pressure head and $K$ is the hydraulic conductivity of the
  medium.  Conductivity is known to depend on porosity, and for variably
  saturated media both $K$ and $h$ are typically taken to be functions
  of saturation, such as in the van Genuchten or Brooks-Corey
  models~\cite{bear-1988}.  Therefore, Darcy's law takes the form of
  Eq.~{\protect\en{darcy}} with $D=K(\sat,\porosity)\, dh/d\sat$.},
where the \emph{effective diffusivity} $D(\sat,\porosity)$
\units{cm^2/\mday} is a function of both water content and porosity (for
which a specific functional form will be presented in
Section~\ref{sec:moist-coe}).  After substituting \en{darcy} into
\en{sat}, we obtain
\begin{gather}
  \frac{\partial \sat}{\partial t} = \frac{\partial
  }{\partial x}\left[ D(\sat,\porosity) \, \frac{\partial \sat}{\partial x}
  \right] - \stoich \satfac \, \frac{\Mwater\,\rCSH}{\rhowater\,\MCSH}
  \label{eq:satnew}.
\end{gather}

Dissolved alite and belite are advected with the pore liquid as well as
being affected by diffusion and reaction, and so the corresponding
conservation equations are
\begin{align}
  \frac{ \partial \left( \sat \Calite\right)}{\partial t} &=
  \frac{\partial}{\partial x}\left( \sat D_\alite\, \frac{\partial
      \Calite}{\partial x} \right)-\frac{\partial \left( u
      \Calite\right)}{\partial x}- \satfac \, r_\alite,
  \label{eq:alite} \\
  \frac{ \partial \left( \sat \Cbelite\right)}{\partial t} &=
  \frac{\partial}{\partial x}\left( \sat D_\belite \,\frac{\partial
      \Cbelite}{\partial x} \right)-\frac{\partial \left( u
      \Cbelite\right)}{\partial x}- \satfac \, r_{\belite},
  \label{eq:belite}
\end{align}
where $D_\constit$, $\constit=\alite, \belite$ is the diffusivity of
each dissolved constituent.  Transport of aqueous \CSH is governed by
\begin{align}
  \begin{split}
    \frac{ \partial \left( \sat \CCSHaq\right)}{\partial t} =\; &
    \frac{\partial}{\partial x}\left( \sat D_\aq \,\frac{\partial
        \CCSHaq}{\partial x} \right)-\frac{\partial \left( u
        \CCSHaq\right)}{\partial x} \\
    & + \; \satfac \left( \rCSH - \kads \CCSHaq +
      \kdes\CCSHgel \right),
    \label{eq:aqgel}
  \end{split}
\end{align}
where $\kads$ and $\kdes$ are the rates of \CSH precipitation and
dissolution respectively.  The solid \CSH phase is not affected by
advective or diffusive transport and so obeys a simple ODE
\begin{gather}
  \frac{ \partial \left( \sat \CCSHgel\right)}{\partial t} =
  \satfac \left(  \kads \CCSHaq - \kdes \CCSHgel \right) .
  \label{eq:solgel}
\end{gather}

In summary, the governing equations consist of \en{satnew},
\en{alite}--\en{solgel}, which enforce conservation of water, aqueous
species, and solid \CSH gel, supplemented with the relationships
\en{porosity} and \en{darcy}.

\subsubsection{Analogy with bioremediation models}
\label{sec:biofilm}

Before presenting the remaining details, it is worthwhile mentioning
that there is a great deal of similarity between our model for reactive
transport in concrete and those developed for biofilm growth and
bioremediation in the soil sciences literature (for
example,~\cite{clement-hooker-skeen-1996,chapwanya-etal-2008,kildsgaard-engesgaard-2001}. 
In the case of bioremediation, bacteria are employed in porous aquifers
in order to break down some targeted contaminant.  Nutrients (typically
nitrates) are injected into the ground to activate the decontamination
process and soil scientists are interested in understanding how to
encourage the growth of the bacteria in a controlled manner so as to
avoid clogging the pores in the rock or soil matrix while at the same
time maximizing the breakdown of contaminant.  The governing equations
for both problems therefore have a similar structure, with a few key
differences that we summarize below:
\begin{itemize}
\item In biofilms water is an inert phase, whereas in concrete it
  participates in the reaction.
\item Biological organisms are typically modelled using Monod reaction
  terms, whereas we use power-law kinetics.
\item Biofilms are composed of living cells and so give rise to
  additional terms that encompass cell division and death processes.
\item The microstructure of biofilms and \CSH are quite different, but
  our use of a continuum approach means that we can ignore such details.
  We do nonetheless employ the same power-law form \en{dfunction} of the
  permeability correction as that used in biofilms.
\end{itemize}

\subsection{Initial and boundary conditions}
\label{sec:icbc}

We assume that the concrete sample at the beginning of an experiment is
homogeneous in composition and uniformly hydrated so that the initial
water content and concentrations for $0<x<\Xmax$ are
\begin{gather}
  \begin{split}
    \sat(x,0)=\satmin, \qquad
    \Calite(x,0) = \Calite^o, \qquad
    \Cbelite(x,0)= \Cbelite^o, \\
    \CCSHaq(x,0)  = \CCSHaq^o, \qquad
    \CCSHgel(x,0) = \CCSHgel^o, \qquad\qquad
  \end{split}
  \label{eq:ic-all}
\end{gather}
where the zero superscript denotes a constant initial value.  The first
condition on water content states that the concrete is initially at the
minimum value, corresponding for example to a sample that is
equilibrated in a humidified environment but not force-dried.  It is
reasonable to take the initial \CSH concentrations
$\CCSHaq^o=\CCSHgel^o=0$, but the alite and belite concentrations are
key model parameters that depend on the composition of the initial
cement mixture.  In particular,
Papadakis \etal~\cite{papadakis-vayenas-fardis-1989} calculate the
initial concentrations as
\begin{subequations}\label{eq:ic-ab}
  \begin{gather}
    C^o_\constit = (1-f_\constit) \, \omega_\constit \, \Comix
    \label{eq:ic-ab1}
    \\
    \intertext{where}
    \Comix = \frac{\rhocem}{\Rwc \,\rhocem/\rhowater + \Rac \,
      \rhocem/\rhoagg + 1} 
    \label{eq:ic-ab2}
  \end{gather}
\end{subequations}
represents the initial concentration of cement before onset of
hydration, $\rhocem$ is the original cement density, $\rhoagg$ is the
particle density of aggregates, $\Rwc$ and $\Rac$ are initial
water-to-cement and aggregate-to-cement ratios by mass, and
$\omega_\constit$ is the mass fraction for each constituent
$\constit=\alite,\belite$.  We have modified Papadakis \etal's formula
slightly to include the extra factors $(1-f_\constit)$ where
$f_\constit\in[0,1]$ represents the fractional degree of hydration of
each constituent at the end of the hydration/curing stages.


The cement mixtures investigated in~\cite{barrita-etal-2004} contain
significant levels of tricalcium aluminate (or \CthreeA, short for
$3\CaO\cdot\AlO$) and no tetracalcium aluminoferrite.  Consequently, for
the purposes of calculating initial porosity, we also include the effect
of \CthreeA, whose initial hydration products further reduce the pore
space available for transport. Letting $f_\gamma$ and $\omega_\gamma$
refer to the mass and hydration fractions for \CthreeA, we are led to
the following expression for initial
porosity~\cite[Tab.~2]{papadakis-vayenas-fardis-1989}:
\begin{gather}
  \porref = \Comix \Rwc/\rhowater - \Comix \left(
    f_\alite  \omega_\alite  \,\Delta V_\alite  +
    f_\belite \omega_\belite \,\Delta V_\belite +
    f_\gamma  \omega_\gamma  \,\Delta V_\gamma
  \right) ,
  \label{eq:ic-gel}
\end{gather}
where the first term represents the porosity before onset of hydration
and the remaining terms encompass the reduction in pore volume owing to
hydration through parameters $\Delta V_\alite=0.233$, $\Delta
V_\belite=0.228$ and $\Delta V_\gamma=0.555$ (units of \units{cm^3/g}).

We note in passing that the strength of the resulting hardened concrete
is related to $\Rwc$ and $\Rac$ as well as the curing conditions.  For
example, a high value of $\Rwc$ yields a low strength concrete owing to
an increase in porosity that occurs because of the excess water present
during hydration; consequently, most concrete is mixed with an initial
water-to-cement ratio ranging from 0.30 to 0.60.

The bottom end of the concrete sample is immersed in water, where we
impose the following Dirichlet boundary condition
\begin{gather}
  \sat(0,t) = \satmax - \frac{\CCSHgel(0,t)}{\rhoCSHgel},
  \label{eq:bc-sat0}
\end{gather}
which states simply that the sample is fully saturated at $x=0$.
We also assume perfect sink conditions on the aqueous species,
so that
\begin{gather}
  C_\constit(0,t)=0 \quad \text{for $j=\alite, \belite, \aq$.}
  \label{eq:bc-abq0}
\end{gather}


In typical experiments, the concrete sample is coated on the sides and
top face with a sealant (such as epoxy) that prevents any transport
into or out of the sample.  This supports our 1D approximation and
allows us to impose homogeneous Neumann boundary conditions
\begin{gather}
  \frac{\partial \sat}{\partial x}(\Xmax,t)=0
  \qquad \text{and} \qquad
  \frac{\partial C_\constit}{\partial x}(\Xmax,t)=0,
  \label{eq:bc-allL}
\end{gather}
where $\constit=\alite,\belite,\aq$.  These conditions are equivalent to
imposing a zero flux because the boundary condition on $\sat$ at
$x=\Xmax$ requires that the convective flux component is zero.  We note
in closing that no boundary conditions are needed for $\CCSHgel$
because it is governed by an ODE.


\subsection{Reaction rates}
\label{sec:rates}

The reaction terms are specified using notation introduced by
Papadakis \etal~\cite{papadakis-vayenas-fardis-1989} wherein the rate of
generation $r_\constit$ \units{g/cm^3\, \mday} of species
$\constit=\alite,\belite$ is
\begin{gather}
  r_\constit = k_\constit C_\constit\, \left( \frac{C_\constit}{C^o_\constit}
  \right)^{n_\constit-1},
  \label{eq:rates}
\end{gather}
with $k_\constit$ \units{\mday^{-1}} a rate constant, $n_\constit$ a
power-law exponent, and $C^o_\constit$ the initial concentration (all
given in Table~\ref{tab:params}).  The total rate of generation of $\CSH$
due to the alite and belite reactions \units{g/cm^3\,\mday} is
\begin{gather}
  \rCSH = \frac{\MCSH}{2} \left( \frac{r_\alite}{m_\alite} +
    \frac{r_\belite}{m_\belite} \right),
  \label{eq:Rcsh}
\end{gather}
where $m_\alite$, $m_\belite$ and $\MCSH$ are molar masses of alite,
belite and \CSH respectively.  A power-law reaction mechanism
similar to \en{rates} has also been employed in other models of cement
chemistry~\cite{meier-etal-2007,saetta-etal-1995,delmi-etal-2006}.

\subsection{Moisture diffusion coefficient}
\label{sec:moist-coe}

Following the approach used for biofilms
in~\cite{clement-hooker-skeen-1996} we take the effective diffusivity to
be a separable function of the form
\begin{gather}
  D(\sat,\porosity) = \varphi^{19/6} D^*(\sat) ,
  \label{eq:dfunction}
\end{gather}
where the influence of porosity on clogging appears as a power law in
the quantity $\varphi=\frac{\porosity-\satmin}{\porref-\satmin}$.  
Clement \etal\ initially assume that the ratio of porosities obeys
$\varphi=(R/R_o)^m$, where $R$ and $R_o$ represent the
corresponding pore radii and $m$ is an empirical constant.  They then
take two very common functional forms of the constitutive laws for
porous media (namely, the van Genuchten and Brooks-Corey relationships)
and show that the hydraulic conductivity in both cases satisfies
$K/K_o=\varphi^{(5m+4)/2m}$; the diffusivity must obey
a similar relationship since it is proportional to $K$.  By comparing
with experimental data from a wide range of soils,
Clement \etal\ find their power-law fit to be insensitive to
the specific choice of $m$.  They conclude that $m=3$ is a reasonable
approximation, which corresponds to the exponent $19/6$ used above.
  
The question remains whether these relationships applied successfully to
biofilm growth in soils are also applicable to \CSH gel formation in
concrete.  It is certainly true that the physics governing the two
processes are very different.  Nonetheless, models for formation of \CSH
are based on the premise that the gel precipitates as outgrowths from
the surface of cement grains~\cite{birchall-howard-bailey-1978} which is
analogous to the manner in which biofilms accumulate on soil particles.
Furthermore, the derivation above uses only spatially averaged
quantities and hence makes no assumption about any specific pattern of
biofilm growth.  We therefore conclude that the $19/6$ rule should also
be applicable to concrete.

Turning now to $D^*(\sat)$, we observe that many other studies of water
transport in concrete and related porous
media~\cite{bary-sellier-2004,hall-hoff-skeldon-1983,pel-1995,mainguy-etal-2000}
approximate the diffusivity by an exponential function of saturation
\begin{gather}
  D^*(\sat)=A e^{B\sat},
  \label{eq:dexp}
\end{gather}
where parameters $A$ \units{cm^2/\mday} and $B$ are empirical constants.
Lockington \etal~\cite{lockington-parlange-dux-1999} performed extensive
experiments which showed that a number of building materials may be
characterized by a universal exponent represented by the rescaled
parameter $\overline{B}=B\, (\satmax-\satmin)$ whose value lies between
4 and 6; other work~\cite{pel-1995} suggests that $\overline{B}$ could
be as low as 2 and as high as 8.  Note that these parameters lead to
very rapid variations in diffusivity over the physical range of
saturations (by at least three orders of magnitude) which distinguishes
water transport in concrete from that of many other common porous media.


\subsection{Choice of base case parameters}
\label{sec:base-params}

The numerical simulations in this paper focus on reproducing
experimental results reported by
Barrita \etal~\cite{barrita-2002,barrita-etal-2004} and specifically the
concrete sample they refer to as ``mixture 3.''  We begin by selecting a
representative set of parameters for a ``base case'' simulation, but
since not all of the required data is provided in these references we have
had to estimate certain values using other literature sources.  The
parameters are summarized in Table~\ref{tab:params} and we comment below
on a number of the more critical choices:
\newcommand{\myditto}{\mbox{\hspace*{0.5cm}{\tt "}}}
\begin{table}[tbhp]
  \caption{Parameter values corresponding to the base case.}
  \label{tab:params}
  \begin{tabular}{cm{0.40\textwidth}cc>{\rrcol}m{0.30\textwidth}}\hline
    Symbol      & Description              & Value    & Units      & Reference\\ \hline
    $\rhowater$ & Liquid water density     &      1.0 & $g/cm^3$   & \\
    $\rhoCSHgel$& \mbox{}\CSH gel density  &      2.6 & $g/cm^3$   & Allen~\etal~\cite{allen-thomas-jennings-2007}\\
    $\rhocem$   & Cement density           &     2.83 & $g/cm^3$   & Barrita~\etal~\cite{barrita-etal-2004} \\
    $\rhoagg$   & Aggregate particle density&     2.6 & $g/cm^3$   & \\
    $m_\alite$  & Alite molar mass         &    228.3 & $g/mol$    & \\
    $m_\belite$ & Belite molar mass        &    172.2 & $g/mol$    & \\
    $\Mwater$   & Water molar mass         &     18.0 & $g/mol$    & \\
    $\MCSH$     & \mbox{}\CSH molar mass   &    342.4 & $g/mol$    & \\
    $D_\alite$  & Alite diffusivity        &     0.01 & $cm^2/\mday$& \\
    $D_\belite$ & Belite diffusivity       &     0.01 & $cm^2/\mday$& \\
    $D_\aq$     & \mbox{}\CSH (aq) diffusivity&  0.01 & $cm^2/\mday$& \\
    $A$         & Water diffusion coefficient&0.0028&$cm^2/\mday$& \\
    $B$         & Water diffusion exponent &    100 & $-$         & Lockington~\etal~\cite{lockington-parlange-dux-1999}\\
    $\satmin$   & Residual water content   &     0.04 & $-$        & Barrita~\cite{barrita-2002} \\
    $\satrx$    & Reaction cut-off         &     0.04 & $-$        & Equal to $\satmin$ \\
    $k_\alite$  & Alite reaction rate      &     22.2 &$\mday^{-1}$& Papadakis~\etal~\cite{papadakis-vayenas-fardis-1989}\\
    $k_\belite$ & Belite reaction rate     &     3.04 &$\mday^{-1}$& \myditto \\
    $n_\alite$  & Alite reaction exponent  &     2.65 & $-$        & \myditto \\
    $n_\belite$ & Belite reaction exponent &     3.10 & $-$        & \myditto \\
    $\kads$     & \mbox{}\CSH precipitation rate&32.2 &$\mday^{-1}$& Bentz~\cite{bentz-2006}\\
    $\kdes$     & \mbox{}\CSH dissolution rate&     0 &$\mday^{-1}$& \myditto \\
    $\stoich$   & Water stoichiometry      &        5 & $-$        & {\protect Eqs.~\en{rx-alite} and \en{rx-belite}}\\
    $\Rwc$      & Water-to-cement ratio    &    0.333 & $-$        & Barrita~\etal~\cite{barrita-etal-2004}\\
    $\Rac$      & Aggregate-to-cement ratio&     2.86 & $-$        & \myditto \\
    $\omega_\alite$ & Alite mass fraction  &     0.65 & $-$        & \myditto \\
    $\omega_\belite$& Belite mass fraction &     0.17 & $-$        & \myditto \\
    $\omega_\gamma$ & \CthreeA\ mass fraction &  0.11 & $-$        & \myditto \\
    $f_\alite$  & Alite hydration fraction &     0.60 & $-$        & Tennis\ \&\ Jennings~\cite{tennis-jennings-2000}\\
    $f_\belite$ & Belite hydration fraction&     0.20 & $-$        & \myditto \\
    $f_\gamma$  & \CthreeA\ hydration fraction&  0.72 & $-$        & \myditto \\
    $\Delta V_\alite$ & Alite volume change&    0.233 & $cm^3/g$   & \myditto \\
    $\Delta V_\belite$& Belite volume change&   0.228 & $cm^3/g$   & \myditto \\
    $\Delta V_\gamma$& \CthreeA\ volume change& 0.555 & $cm^3/g$   & \myditto \\
    $\Xmax$     & Length of sample         &     10.0 & $cm$       & Barrita~\etal~\cite{barrita-etal-2004} \\
    $\CCSHaq^o$ & Initial \CSH (aq) concentration & 0 & $g/cm^3$   & \\
    $\CCSHgel^o$& Initial \CSH (gel) concentration& 0 & $g/cm^3$   & \\
    \hline
    \multicolumn{5}{l}{\emph{Derived parameters:}} \\
    $\Calite^o$ & Initial alite concentration & 0.145 & $g/cm^3$   & Eq.~{\protect\en{ic-ab}}\\
    $\Cbelite^o$& Initial belite concentration& 0.076 & $g/cm^3$   & \myditto \\
    $\porref$   & Initial porosity            & 0.067 & $-$        & Eq.~{\protect\en{ic-gel}}\\
    $\satmax$   & Maximum water content       & 0.067 & $-$        & \myditto \\
    \hline
  \end{tabular}
\end{table}

\myparagraph{Sample geometry} 
We have taken the model domain to have length $L=10\;cm$ which is
consistent with the cylindrical samples of concrete used
in~\cite{barrita-etal-2004}.

\myparagraph{Water transport coefficients} 
The maximum water content is $\satmax=0.067$, which is equal to the
initial porosity for the base case and is consistent with measured
values reported in the concrete
literature~\cite{bary-sellier-2004,kumar-bhattacharjee-2003}.  The
residual water content is taken to be a small positive number because
concrete is typically not totally free of water unless it has been
artificially dried~\cite{ferretti-bazant-2006} and in practice some
small amount of water is typically trapped within the porous concrete
matrix; specifically, we choose a value of $\satmin=0.04$ by estimating
the minimum water content from plots in~\cite{barrita-2002}.  We take
the diffusion parameter $B=100$, which is chosen so that the rescaled
quantity $\overline{B}=B\, (\satmax-\satmin)=2.66$ (for the base case
and other simulations performed later) is consistent with the
range of values mentioned in Section~\ref{sec:moist-coe}.  The value of
$A=0.0028$ then follows by fitting the simulated wetting curves to
Barrita's experimental results (more details are provided in
Section~\ref{sec:base-case}).


\myparagraph{Diffusion coefficients for aqueous species}
Since the alite and belite actually dissociate and diffuse as ions, the
best we can do is to use an approximation that represents the
diffusivities in some averaged sense. We begin with the diffusivities of
the ionic constituents $\Ca$, $\OH$ and $\HSO$, which are equal to 0.68,
4.6, and 0.43 $cm^2/d$ respectively~\cite{tzschichholz-zanni-2001}, and
compute an appropriately-scaled harmonic average of approximately
$1.0\;cm^2/d$ for both alite and belite (following the development
in~\cite{farrington-alberty-1966}).  The diffusion of ions in
cementitious materials is known to be reduced by a factor ranging from
$10^{-1}$ to $10^{-3}$~\cite{garboczi-bentz-1992} that depends on the
pore structure and cement composition; in the absence of any better
information we choose a factor of $10^{-2}$ after which
$D_\alite=D_\belite=0.01\;cm^2/d$.  The \CSH gel does not diffuse in
ionic form, and since no data is available in the literature regarding
its diffusion coefficient we have chosen to simply take the same value
$D_q=0.01\;cm^2/d$.  This is not so much of a concern, since we
investigate later on in Section~\ref{sec:vary-Dabq} the effect of
varying $D_\constit$ and demonstrate that the solution is relatively
insensitive to the values of diffusivity.
  
\myparagraph{\CSH composition}
\CSH takes on a whole range of possible forms represented by the general
formula C$_y$S$_2$H$_z$ and so can only be considered in an averaged
sense. We take $\MCSH=342.4\;g/mol$ as a representative molar mass
corresponding to $y=z=3$, which is consistent with many other studies.
There is a correspondingly wide range of gel densities reported in the
literature, from 1.85 $g/cm^3$ at the lower
end~\cite{jennings-etal-2007} up to 3.42
$g/cm^3$~\cite{preece-billingham-king-2001}; we have chosen an
intermediate value of $\rhoCSHgel=2.6\;g/cm^3$ which is justified by
recent work on \CSH microstructure~\cite{allen-thomas-jennings-2007}.

\myparagraph{Cement composition} 
According to information provided in~\cite{barrita-etal-2004} on
concrete mixture 3, the mass fractions of silicate constituents in the
cement are $\omega_\alite=0.65$, $\omega_\belite=0.17$ and
$\omega_\gamma=0.11$, while the aggregate- and water-to-cement ratios
are $\Rac=2.86$ and $\Rwc=0.333$.  The cement mixture also contains
30\%\ by weight of fly ash, which is a lower-density pozzolanic additive
that improves the strength and workability of the resulting concrete.
Based on densities of $3.15$ and $2.08\; g/cm^3$ for Portland cement and
fly ash respectively, this translates into an overall cement density of
$\rhocem=2.86\;g/cm^3$.  All concrete samples were moist cured for 7
days which allows us to estimate $f_\alite=0.60$, $f_\belite=0.20$ and
$f_\gamma=0.72$ from the plot of hydration fractions versus curing time
given in~\cite{tennis-jennings-2000}.  Finally, the aggregates used in
all mixtures are a combination of both fine and coarse quartz materials,
and so we take $\rhoagg=2.6\;g/cm^3$ which is representative of the dry
particle density of sand.

\myparagraph{Alite and belite reaction rates} 
There is considerable variation in rate parameters reported in the
literature owing partly to the fact that many experiments are performed
not on cement samples but rather under idealized equilibrium conditions
in which reactants are in solution.  We have therefore chosen our
parameters based on the data provided
in~\cite{papadakis-vayenas-fardis-1989}, who proposed the mechanism
\en{rates} along with reaction exponents $n_\alite=2.65$ and
$n_\belite=3.10$; however if we use their values of $k_\alite=1.01$ and
$k_\belite=0.138$, then our model exhibits negligible clogging.  But in
fact, the reaction rate coefficients reported in the literature vary by
several orders of
magnitude~\cite{tzschichholz-zanni-2001,bentz-2006,thomas-jennings-1999}
and so this ambiguity has led us to use the reaction rates as fitting
parameters.  Specifically, we take $k_\alite=22.2$ and $k_\belite=3.04$,
which lie within the range of published values while also maintaining
the same ratio of $k_\alite/k_\belite$ used
in~\cite{papadakis-vayenas-fardis-1989} (more details on the fitting
procedure are provided in Section~\ref{sec:base-case}).
  
\myparagraph{Precipitation and dissolution rates}
Bentz~\cite{bentz-2006} developed a model that assumes a linear
hydration rate law with rate constant ranging from 0.264 to
$1.464\;\mday^{-1}$ depending on $\Rwc$.  We choose the precipitation
rate as the upper end of their range, but again scale using the same
factor as the other reaction rates to obtain $\kads=32.2$.  We also take
$\kdes=0$ following Bentz and others who neglect \CSH dissolution.

\section{Numerical simulations}
\label{sec:numerics}

The governing equations are discretized in space using a centered finite
volume approach wherein the domain is divided into $N$ uniform cells
having width $\dx=\Xmax/N$ and centered at $x_i= (i-1/2)\dx$ for $i=1,
2, \dots, N$.  The discrete solution components, for example
$C_i(t)\approx \Calite(x_i,t)$, are approximations of the average value
of the solution within each cell.  Using this notation, the discrete
approximation of the alite equation~\en{alite} is
\begin{align}
  \begin{split}
    \frac{ \partial (\sat_i C_i)}{\partial t}
     = \; & \frac{D_\alite}{\dx}\left( \sat_{i+1/2} \,
      \frac{C_{i+1} - C_{i}}{\dx}
      - \; \sat_{i-1/2} \, \frac{C_{i}-C_{i-1}}{\dx}\right)
    \\
    & - \frac{u_{i+1/2} C_{i+1/2} - u_{i-1/2} C_{i-1/2}}{\dx}
    - (\sat_i-\satrx)^+ ({r_\alite})_i,
  \end{split}
  \label{eq:discrete}
\end{align}
where the quantities $C_{i\pm 1/2}$ are approximations of the solution
at the left ($-$) and right ($+$) cell edges for which we use an
arithmetic average $C_{i\pm 1/2}=(C_i+C_{i\pm 1})/2$.  The discrete
velocity at cell edges is written using the centered difference
approximation of Darcy's law
\begin{gather*}
  u_{i-1/2} = - D(\sat_{i-1/2}, {\porosity}_{,i-1/2}) \,
  \frac{ \sat_{i} - \sat_{i-1} }{\dx}.
\end{gather*}
The same approach is used to discretize the remaining conservation
equations \en{satnew}, \en{belite}, \en{aqgel} and \en{solgel}.  In all
cases, the equations corresponding to boundary cells $i=1$ and $N$
involve ``fictitious'' solution values located at points $x_{0}=-\dx/2$
and $x_{N+1}=\Xmax+\dx/2$ which lie one-half grid cell outside the
domain.  The boundary conditions are discretized using second-order
differences or averages, and are used to eliminate these fictitious
values in terms of interior solution components.

The resulting semi-discretization is fully second order accurate in
space and leads to a system of $5N$ ordinary differential equations for
the discrete solution values which we then integrate in time using
{\scshape Matlab}'s stiff solver {\ttfamily ode15s}.  For all
simulations, we use $N=100$ cells and set both relative and absolute
error tolerances for {\ttfamily ode15s} to $10^{-8}$.  The equations are
integrated to time $t=28$ days, which requires approximately $40\;s$
of clock time on a \mymaclaptop.

\subsection{Base case with and without reactions}
\label{sec:base-case}

We focus on developing comparisons with the experiments of
Barrita~\cite{barrita-2002,barrita-etal-2004} who studied wetting of
concrete cylinders with both water and isopropanol.  The latter solute
is particularly useful in such a study because the silicate compounds in
concrete do not react with isopropanol as they do with water, and so the
isopropanol results may be used to calibrate the diffusion parameter $A$
with experimental data.

\begin{figure}[tbhp]
  \centering
  \tabcapfont
  \myepsfile{runs/front0}{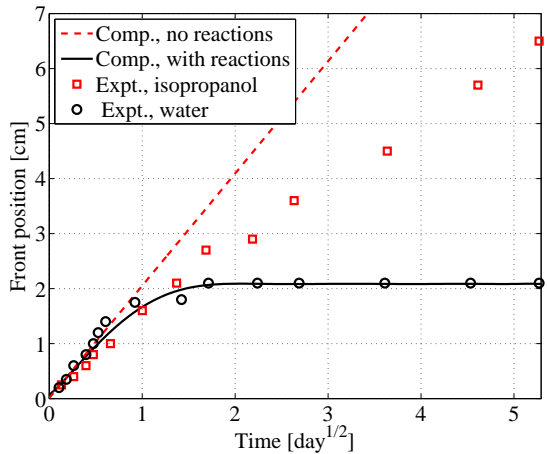}{0.6\textwidth}
  \caption{Wetting front location $s(t)$ for the base case computations
    both with reactions (solid line) and without (dashed line), which
    should be compared to the experimental data with water (circular
    points, taken from~{\protect\cite{barrita-2002}}).  The
    corresponding experimental data for isopropanol (square points) are
    also included for comparison purposes.}
  \label{fig:base-front}
\end{figure}

In the absence of reactions ($k_\alite=k_\belite=\kads=0$) there is no
change in constituent concentrations and so the problem reduces to a
single nonlinear diffusion equation for the water content.  It is well
known that for an exponential diffusivity of the form \en{dexp} with
large $B$ and small $A$, the diffusion equation has a solution which
forms a steep front that progresses into the sample with speed nearly
proportional to the square root of time; consequently, a plot of the
isopropanol wetting front location versus $t^{1/2}$ should be a straight
line, as indicated by the experimental data of Barrita reproduced in
Fig.~\ref{fig:base-front} (square points).  The experimental results for
water uptake also exhibit a linear trend over the first 8--10 hours,
which represents a period over which the reactions have not yet begun to
take hold and no significant clogging has occurred.  However, there is a
noticeable difference between the initial slopes of the isopropanol and
water data, which is most likely attributable to variations in the
concrete samples used, or differences in the capillary pressure or other
transport properties for the two liquids.  Therefore, we have fit our
model to the first 8--10 hours from the water experiment instead of
using the isopropanol data.

We proceed by setting $B=100$ and varying $A$ until the slope of the
wetting front curve best approximates that of the experimental data.
This fitting yields an estimate of $A= 0.0028\;cm^2/\mday$ which is
consistent with values reported by other authors such
as~\cite{akita-etal-1997}.  A plot of the computed wetting front
location (without reactions) is displayed in Fig.~\ref{fig:base-front}
alongside the corresponding experimental data for comparison purposes.
In this and all successive computations, the front location $s(t)$ has
been estimated by identifying the point $x$ where the water content
comes to within some tolerance of $\satmin$; that is, $s(t) = \min \{ x
: \sat(x,t) \leq \satmin + \sattol \}$.\ \ 

\begin{figure}[tbhp]
  \tabcapfont
  \begin{tabular}{c@{\hspace{0.1cm}}c}
    \myepsfile{runs/norx/moist}{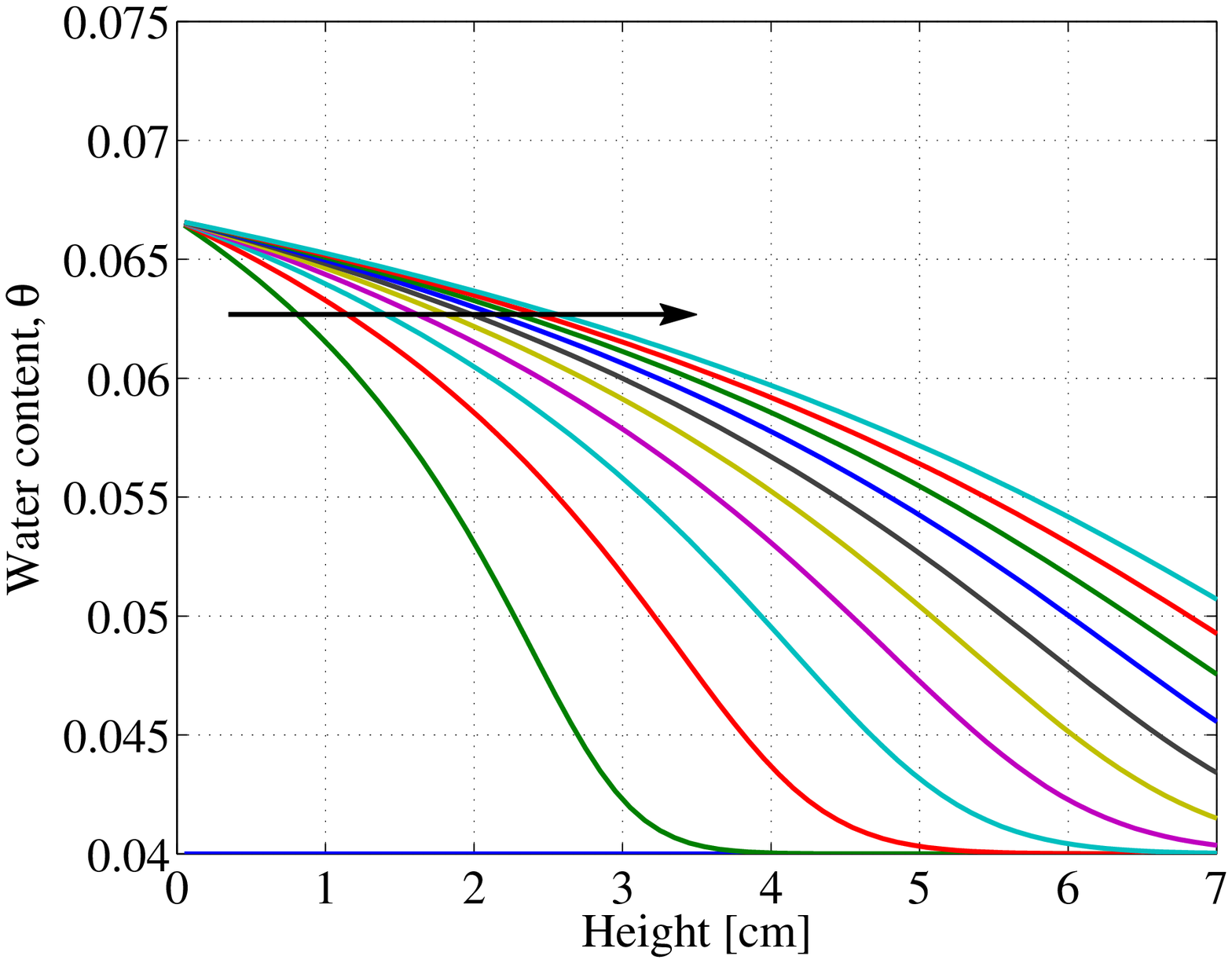}{0.47\textwidth} &
    \myepsfile{runs/base/moist}{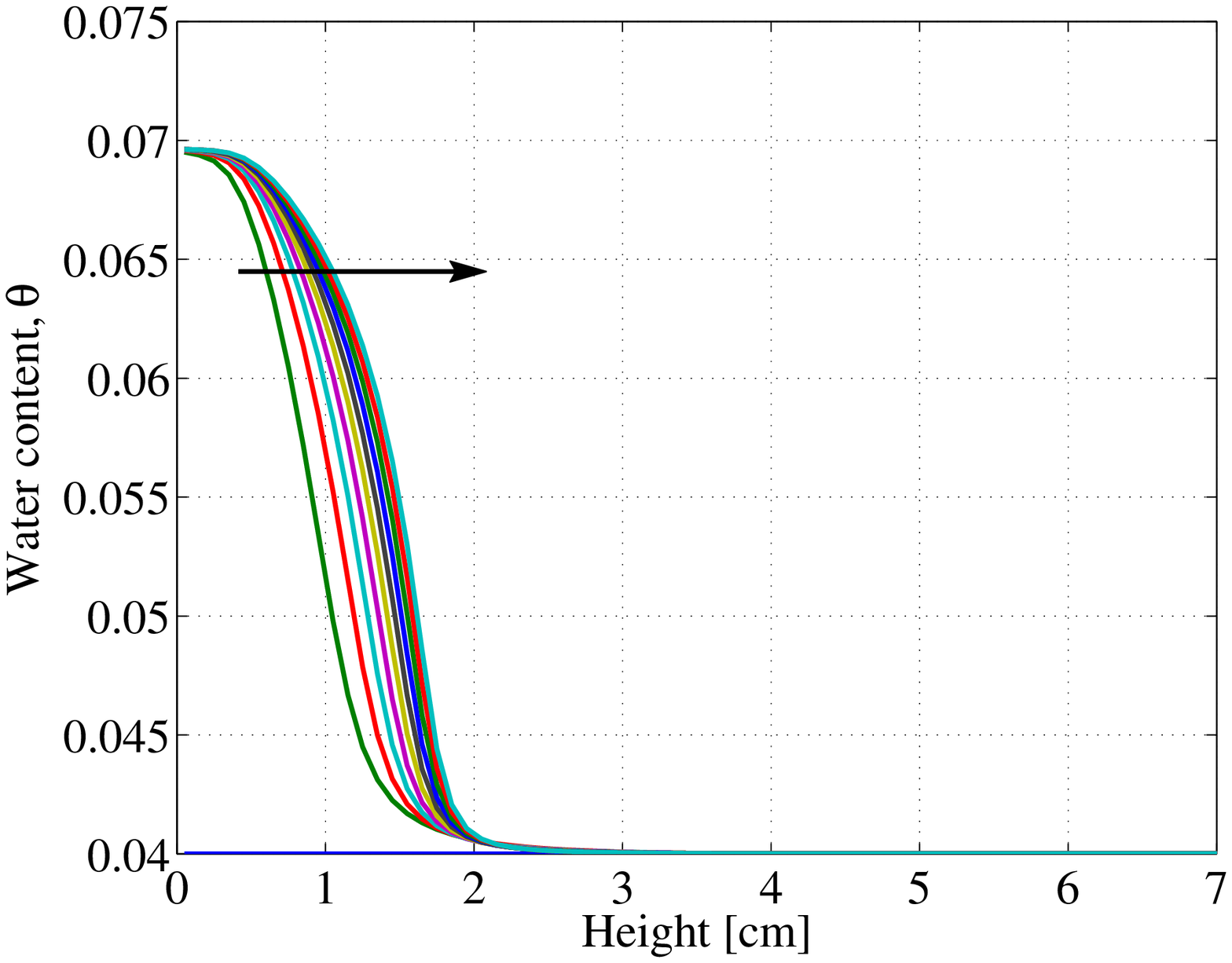}{0.47\textwidth} \\
    a.~~Without reactions. &
    b.~~With reactions. \\
  \end{tabular}
  \caption{Plots of computed water content for the base case parameters,
    both with and without reactions.  The various solution profiles
    correspond to 10 equally-spaced times over the 28~days of the
    simulation.} 
  \label{fig:base-sat}
\end{figure}

Moving now to the case of water uptake including hydration reactions, it
remains to choose an appropriate scaling of rate constants in order to
best match the location of the stalled wetting front in experiments.  As
mentioned earlier in Section~\ref{sec:base-params}, we scale the
reaction and absorption rates $k_\alite$, $k_\belite$ and $\kads$
simultaneously with the same value, while holding their ratio constant
at $1.01 : 0.138 : 1.464$.  This procedure yields the rates
$k_\alite=22.2$, $k_\belite=3.04$ and $\kads=32.2$ for which the
computed water content profile is displayed for comparison purposes in
Fig.~\ref{fig:base-sat}.  We observe that incorporating the effects of
hydration reactions and clogging due to \CSH gel formation clearly
causes the wetting front to stall a short distance inside the
sample. 

\begin{figure}[tbhp]
  \centering
  \tabcapfont
  \begin{tabular}{c@{\hspace{0.1cm}}c}
    \myepsfile{runs/base/alite}{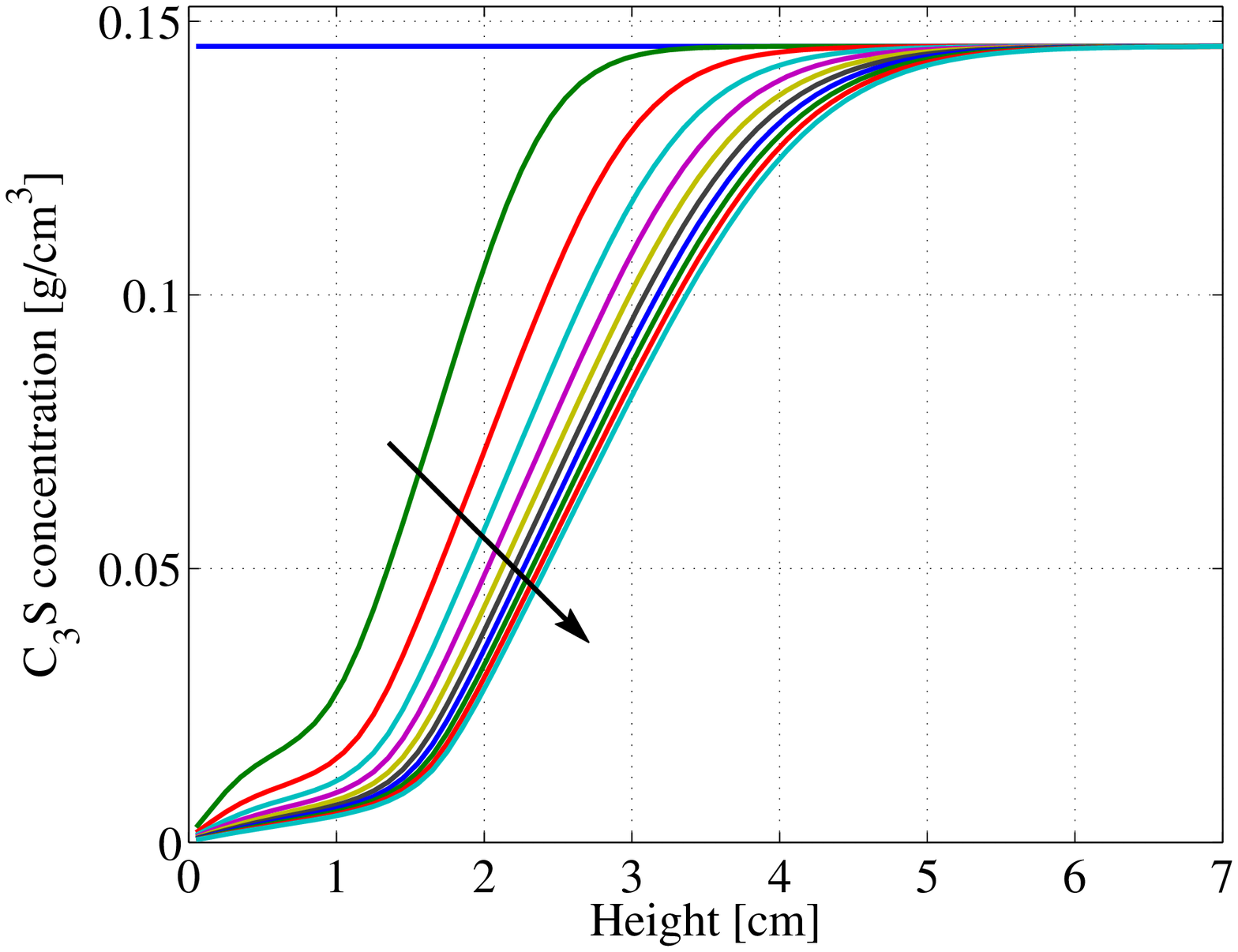}{0.47\textwidth} &
    \myepsfile{runs/base/belite}{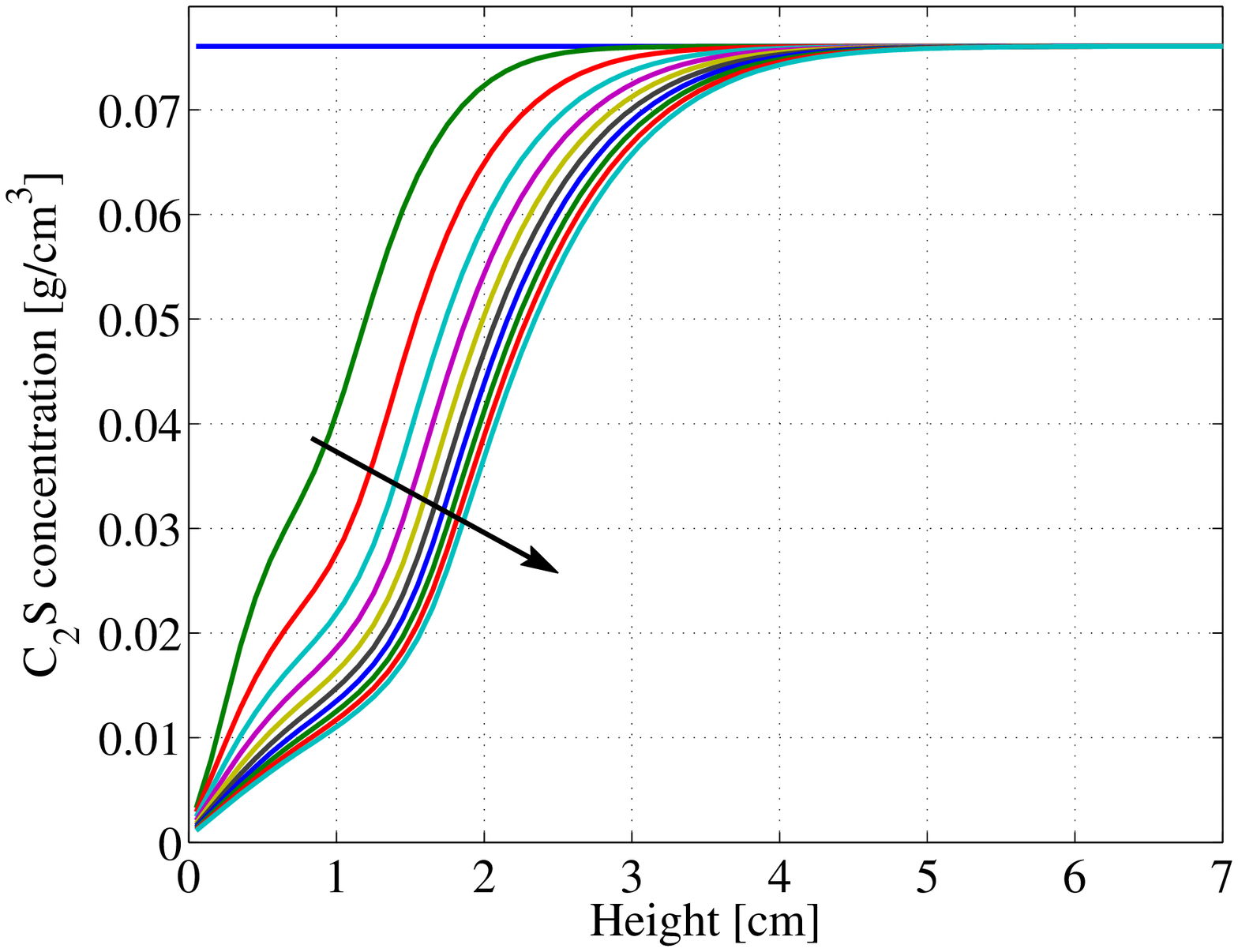}{0.47\textwidth} \\
    a.~~Alite concentration, $\Calite$. &
    b.~~Belite concentration, $\Cbelite$. \\
    \myepsfile{runs/base/csh}{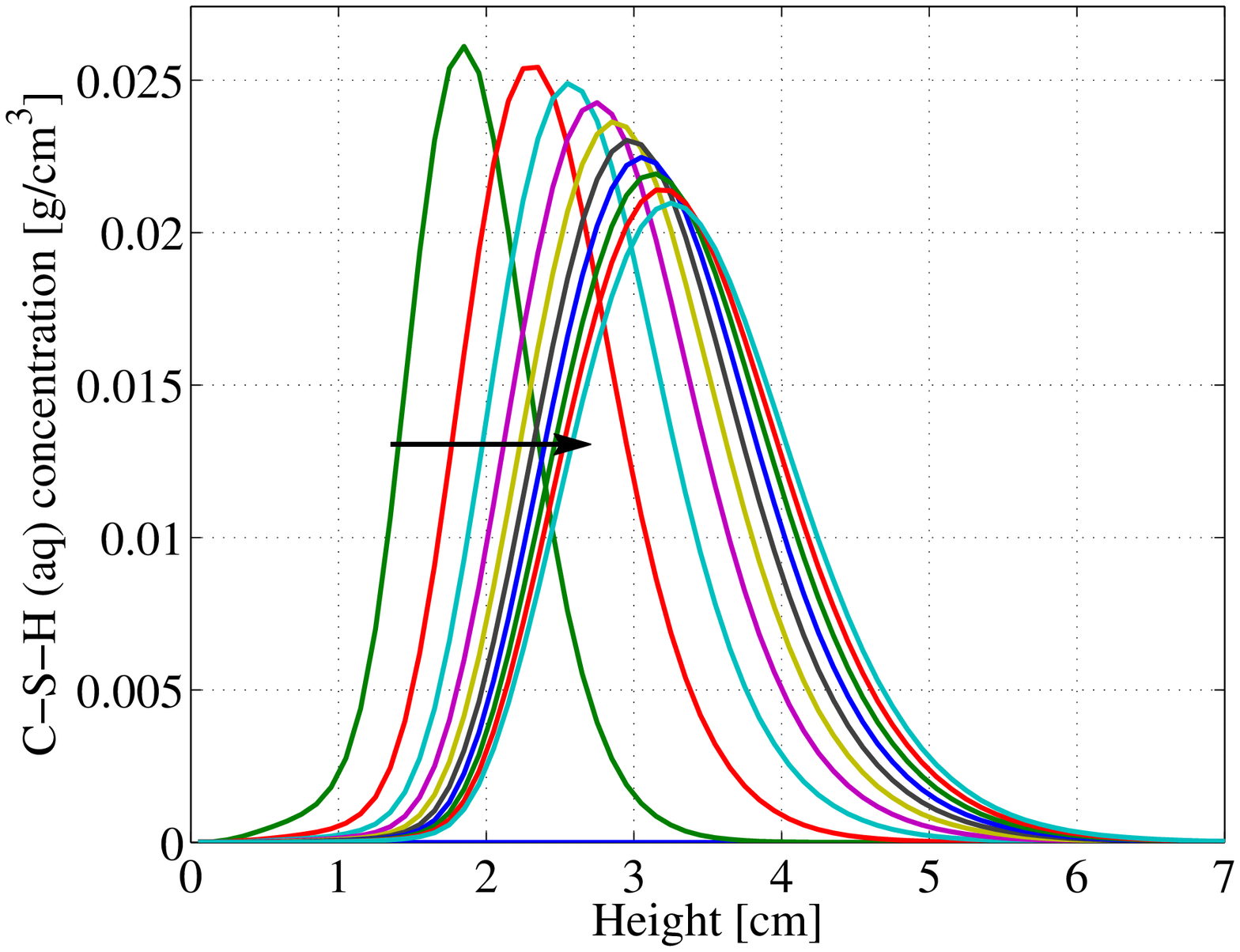}{0.47\textwidth} &
    \myepsfile{runs/base/gel}{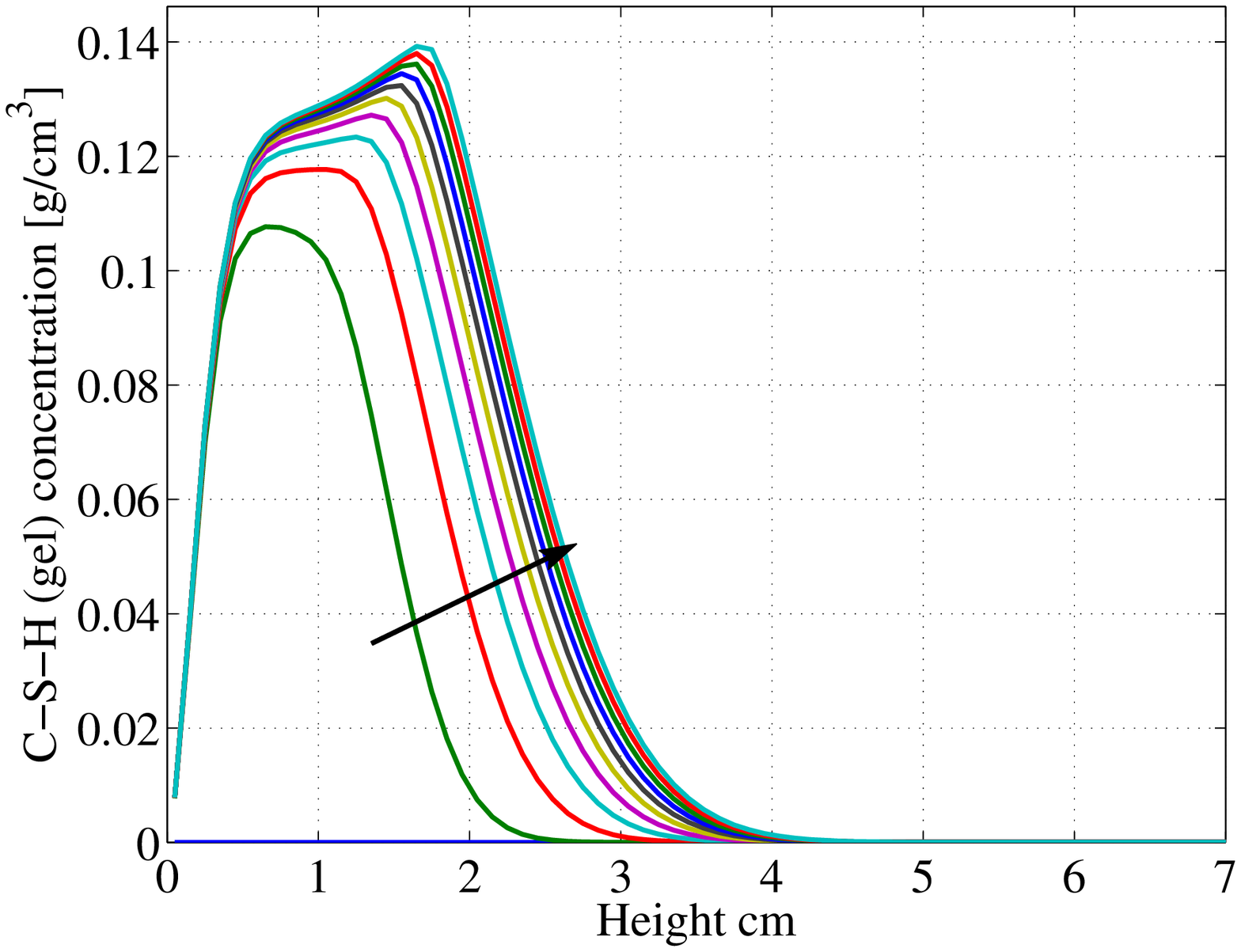}{0.47\textwidth} \\
    c.~~Aqueous \CSH concentration, $\CCSHaq$. &
    d.~~\CSH gel concentration, $\CCSHgel$.\\
  \end{tabular}
  \myepsfile{runs/base/porosity}{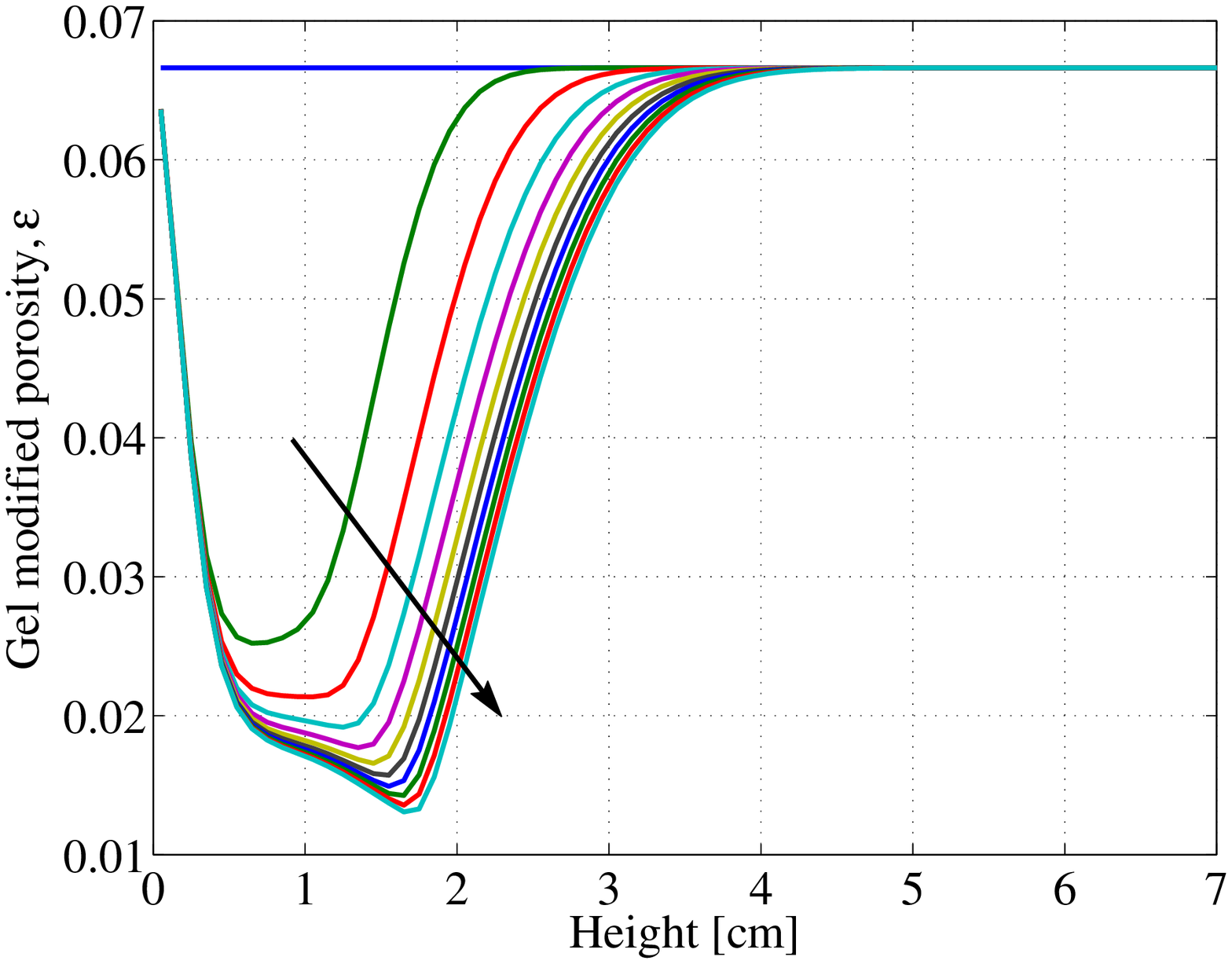}{0.47\textwidth}\\
  e.~~Gel modified porosity, $\porosity$.
  \caption{The remaining base case solution profiles corresponding to
    Figs.~{\protect\ref{fig:base-front}}
    and~{\protect\ref{fig:base-sat}}b.  In each plot, the solution is
    displayed at 10 equally-spaced time intervals over 28~days.  The
    arrows on each plot indicate the progression of curves in the
    direction of increasing time.}
  \label{fig:base-other}
\end{figure}

Plots of concentrations and gel-modified porosity are provided in
Fig.~\ref{fig:base-other}, which indicate how transport of reactants
into the sample is initially dominated by diffusion (for which the front
propagates with velocity proportional to ${t}^{1/2}$) but then later
stalls as \CSH forms and is precipitated near the lower end of the
sample.  The onset of clogging can be clearly seen in the gel
concentration plots where $\CCSHgel$ exhibits a peak slightly behind
the stall location, while the porosity drops to its minimum value
(approx.\ 0.13) within an interval
containing the wetting front and $\CCSHgel$ peak.  It is worthwhile
noting that diffusion and reaction processes continue to occur even
after the front stalls -- most noticeably ahead of the wetting front --
owing to the presence of residual pore water, although this process
continues at a much slower rate.  We emphasize that although the capillary
percolation threshold $\sat=\satmin$ corresponds to the point where
water can no longer move by capillary action, there is still sufficient
water available for the aqueous components to diffuse (since $\sat$
represents the physically bound or absorbed water as well as capillary
water).

The trends shown here suggest that onset of clogging occurs in the
interior of the sample to the right of the inflow boundary.  This effect
can be attributed to a large initial influx of water at $x=0$ that
dissolves the alite and belite near the boundary transporting them some
distance downstream before the gel precipitates.  This result is
consistent with~\cite{barrita-2002} who reported high values of
water flux within the first few hours of their experiments.

We emphasize here that similar stalling behavior has been reported by
several other authors performing experiments on porous building
materials~\cite{taylor-etal-1999,kuntz-lavallee-2001,lockington-parlange-2003}
although these authors attributed this behaviour to an anomalous
diffusion mechanism.  Our primary aim here has been to show that a
similar phenomenon can arise from pore clogging caused by hydration of
residual silicates in concrete.

\begin{figure}[tbhp]
  \tabcapfont
  \begin{tabular}{c@{\hspace{0.1cm}}c}
    \myepsfile{runs/king.D/moist}{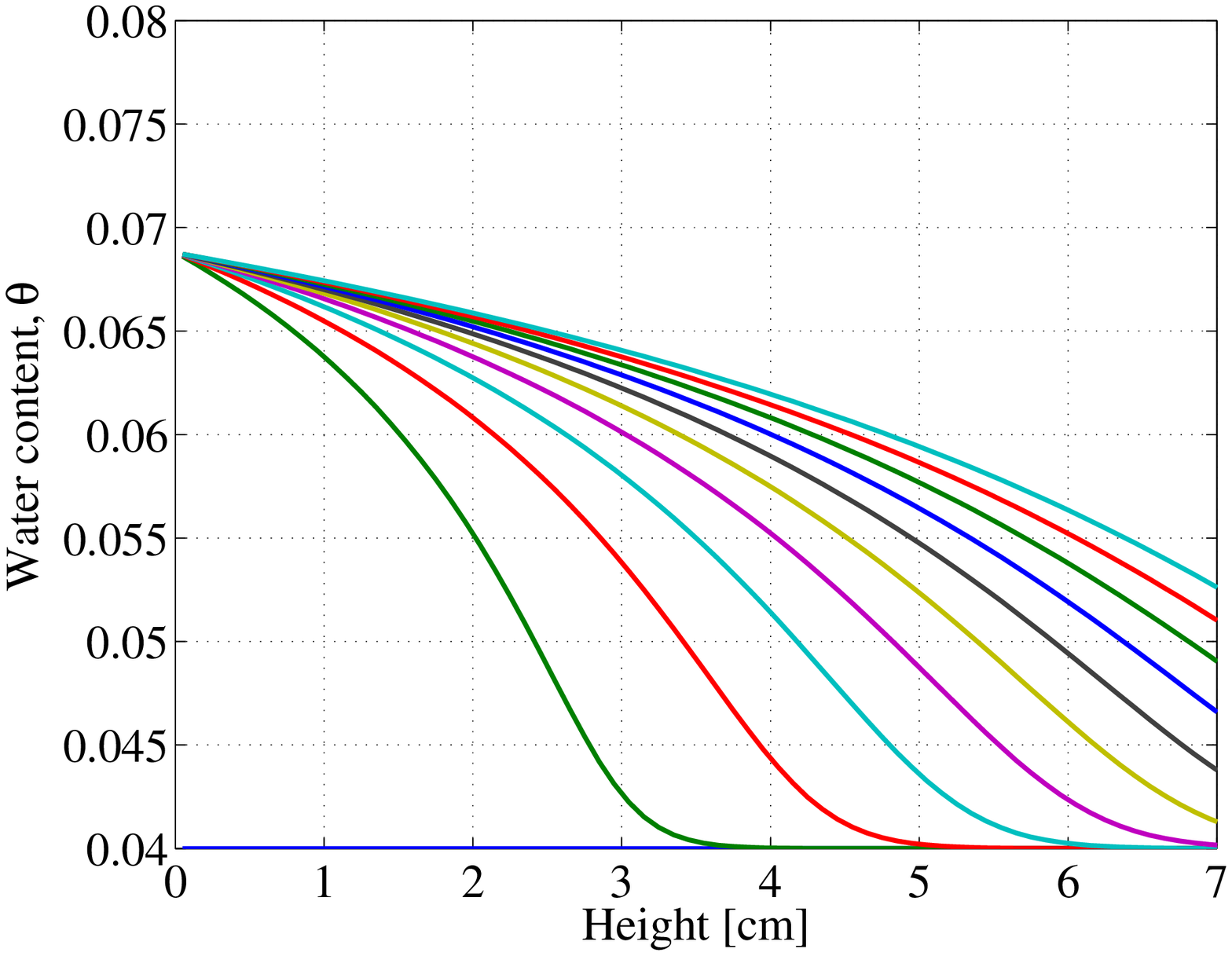}{0.47\textwidth} &
    \myepsfile{runs/king.D/gel}{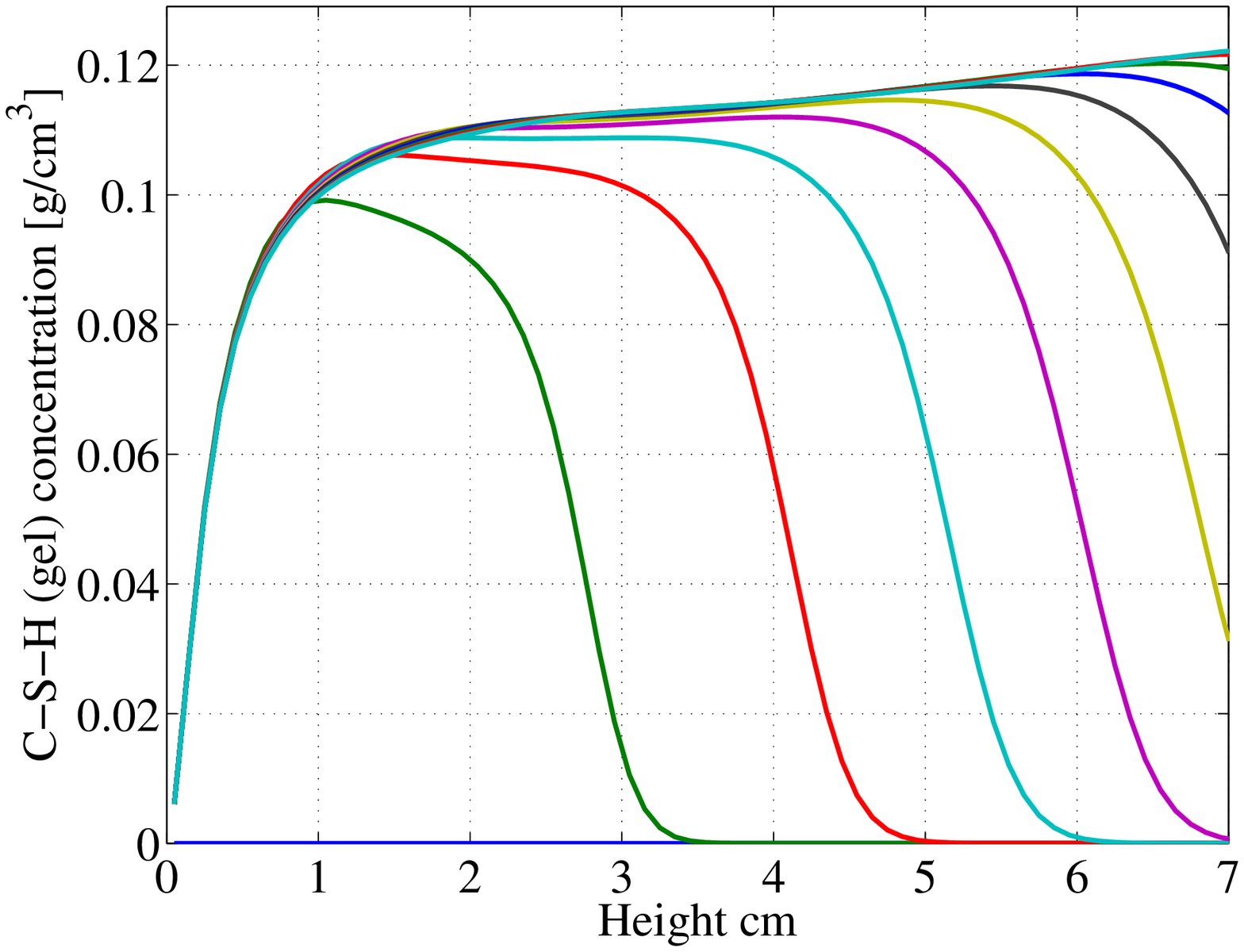}{0.47\textwidth} \\
    a.~~Water content. & 
    b.~~\CSH\ gel concentration. \\
  \end{tabular}
  \caption{Solution with no porosity dependence in the diffusion
    coefficient, exhibiting an absence of clogging.} 
  \label{fig:no-dclogging}
\end{figure}

The formation of a wetting front and subsequent stalling due to pore
clogging are strongly dependent on two components of our model: the
porosity dependence in the diffusion coefficient which drops to zero as
$\porosity\rightarrow \satmin$; and the ''shut-off'' factor $\satfac$
appearing in the reaction terms.  To illustrate the impact of omitting
either effect, we present two additional simulations.  First, if
the porosity correction factor is removed from $D(\sat,\porosity)$ in
Eq.~\en{dfunction}, then the wetting front propagates as if there were
no clogging at all.  This is evident by comparing the plot of water
content in Fig.~\ref{fig:no-dclogging}a with that from the non-reactive
case in Fig.~\ref{fig:base-sat}a.  There is clearly no visible effect on
the front motion, even though a significant level of \CSH\ gel builds up
due to the reactions (see Fig.~\ref{fig:no-dclogging}b).

\begin{figure}[tbhp]
  \tabcapfont
  \begin{tabular}{c@{\hspace{0.1cm}}c}
    \myepsfile{runs/king.rx=5e-5/moist}{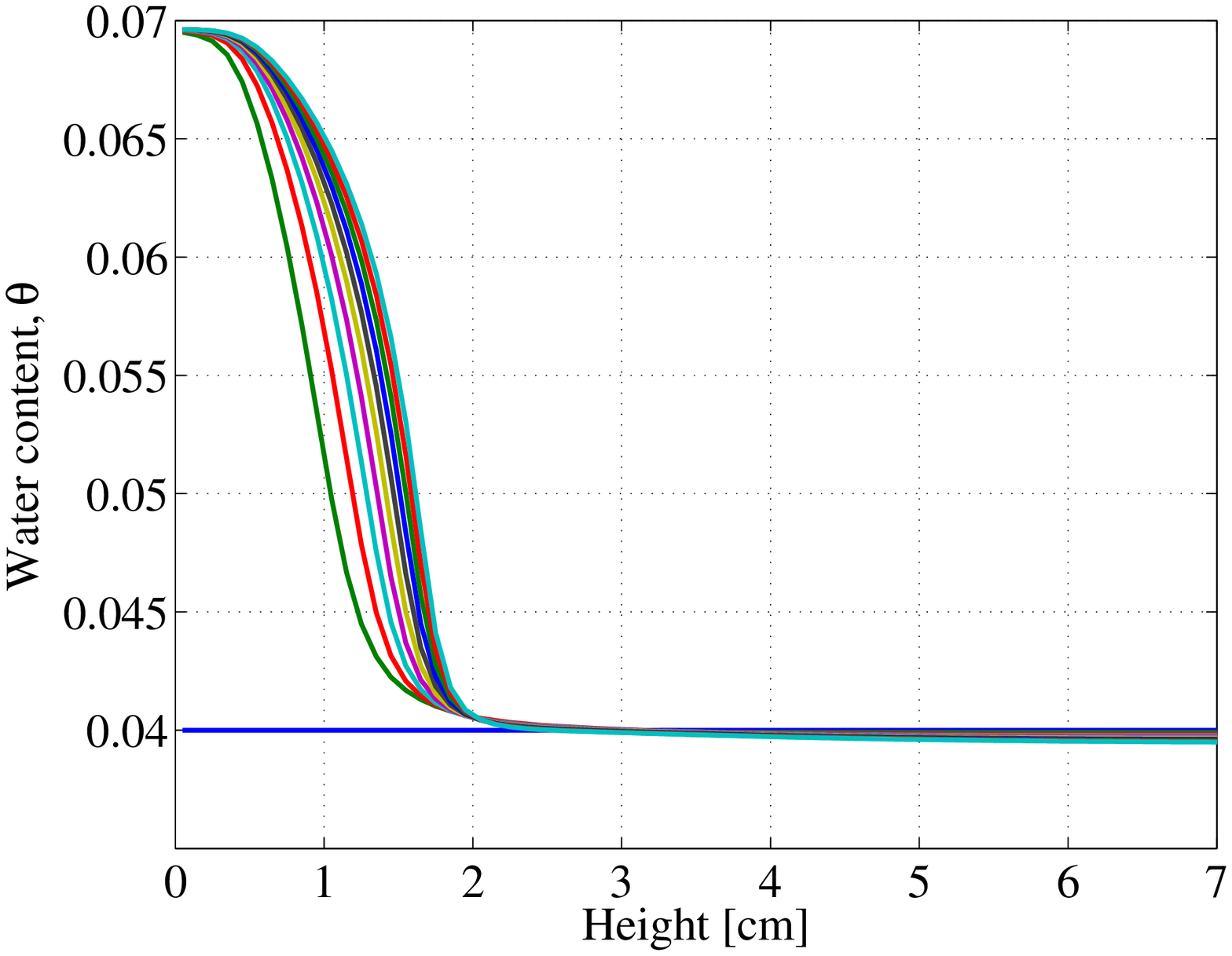}{0.47\textwidth} &
    \myepsfile{runs/king.rx=5e-5/gel}{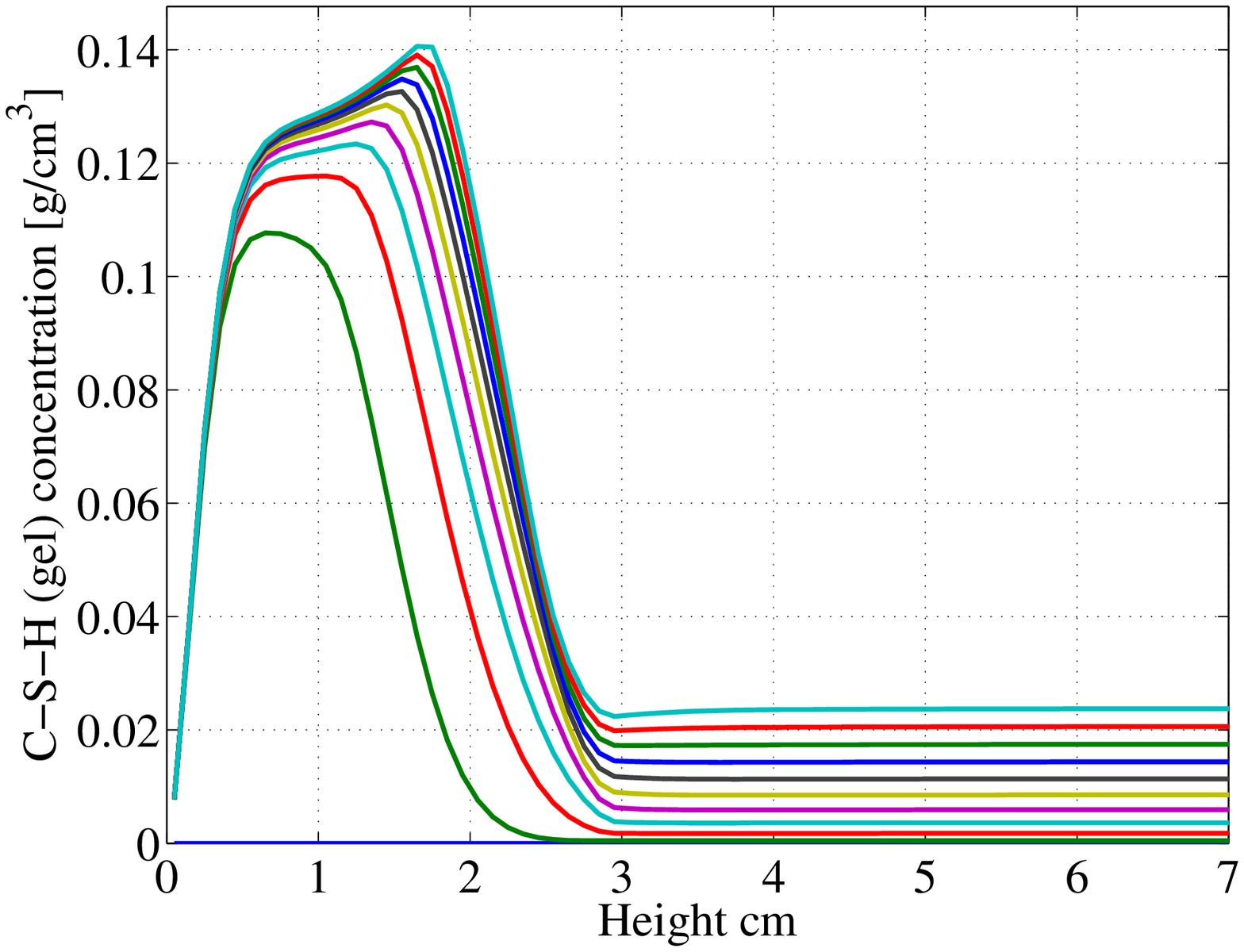}{0.47\textwidth} \\
    a.~~Water content. & 
    b.~~\CSH\ gel concentration. \\
  \end{tabular}
  \caption{Solution computed by replacing
    $\satfac\approx\max(\sat-\satrx,5\times 10^{-5})$, corresponding to
    the situation when reactions do not entirely shut off at the
    residual saturation.}  
  \label{fig:no-shutoff}
\end{figure}

To investigate the effect of slightly relaxing the cut-off factor
$\satfac$ in the reaction terms, we replace the zero cut-off with a
small positive value of $5\times 10^{-5}$ when $\sat \leq \satrx$.  The
wetting front still stalls as indicated in Fig.~\ref{fig:no-shutoff}a;
however, reactions occur over the entire domain giving rise to a
significant concentration of \CSH\ gel to the right of the front and a
corresponding small reduction in saturation below $\satmin$.  This
effect may be attributed to self-desiccation; however, with no more
guidance in how to determine the value of the cut-off parameter, we
leave the study of this effect as a possible avenue for future work and
retain the factor $\satfac$ as is.  Taking a larger value of the cut-off
(close to $\satmin$ in magnitude) can lead to runaway reactions and
instabilities in the numerical method.

We conclude from these last two simulations that in order for our model
to give a reasonable picture of clogging observed in re-wetting
experiments, there must be some retarding of liquid transport through a
porosity dependence in the water diffusivity, and furthermore the reactions
must include a shut-off term similar to $\satfac$, although a small
positive reaction rate might be allowed near the residual saturation.

\subsection{Grid refinement study}
\label{sec:grid-refine}

To ensure that our numerical simulations are computing a consistent
solution that converges with the expected order of accuracy, we
performed a grid refinement study.  The base case simulation was
repeated on successively finer grids with $N=25, 50, 100, 200, 400,
800$, and the solution on the finest grid is treated as the exact
solution.  The solution error was estimated using the discrete $\ell_2$
norm of the difference in aqueous \CSH concentrations $\|\CCSHaq^N -
\CCSHaq^{\text{\emph{finest}}}\|_{\ell_2}$; Any solution component would
suffice, but we choose $\CCSHaq$ because it often displays the greatest
variations.  The results are summarized in Table~\ref{tab:grid}, and the
ratio between successive errors indicates that the solution appears to
be converging at a rate that is at least second order, as expected.

\begin{table}[tbhp]
  \caption{Grid refinement study.  The order is calculated as
    $\log_2(\text{ratio})$.}
  \label{tab:grid}
  \begin{tabular}{cc>{\hspace*{0.5cm}}c>{\hspace*{0.5cm}}c} \hline
    No. of points ($N$) & $\ell_2$-error  & Ratio & Order \\ \hline
    25  & 0.019   & 2.12 & 1.08 \\ 
    50  & 0.0087  & 4.30 & 2.10 \\
    100 & 0.0020  & 5.56 & 2.48 \\
    200 & 0.00036 & 6.27 & 2.65 \\
    400 & 0.00058 & --   & --   \\
    \hline
  \end{tabular}
\end{table}

\subsection{Sensitivity to alite/belite reaction rates}
\label{sec:vary-kab}

\begin{figure}[tbhp]
  \tabcapfont
  \begin{tabular}{c@{\hspace{0.1cm}}c}
    \myepsfile{runs/sumalphsat}{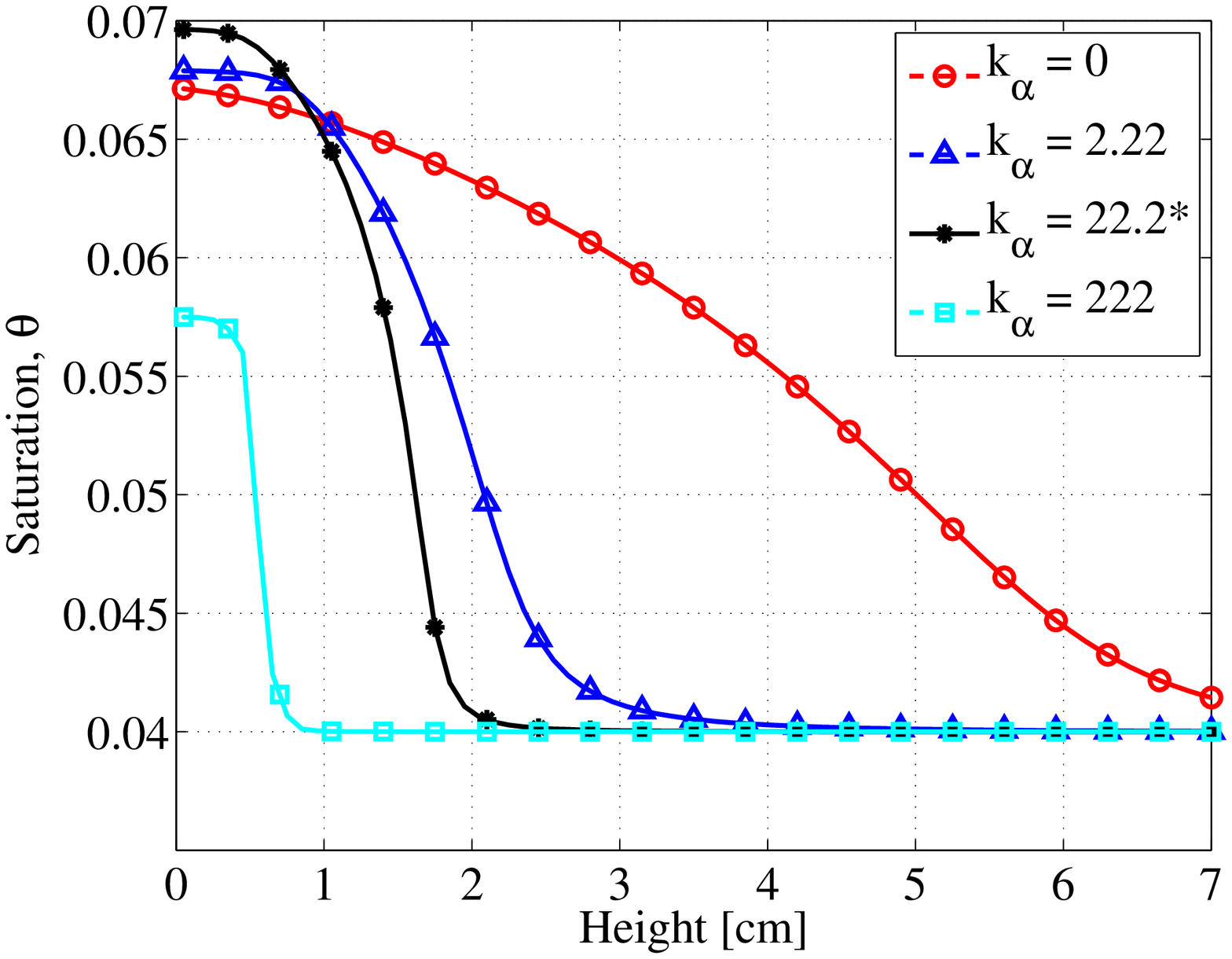}{0.50\textwidth} &
    \myepsfile{runs/sumalphpos}{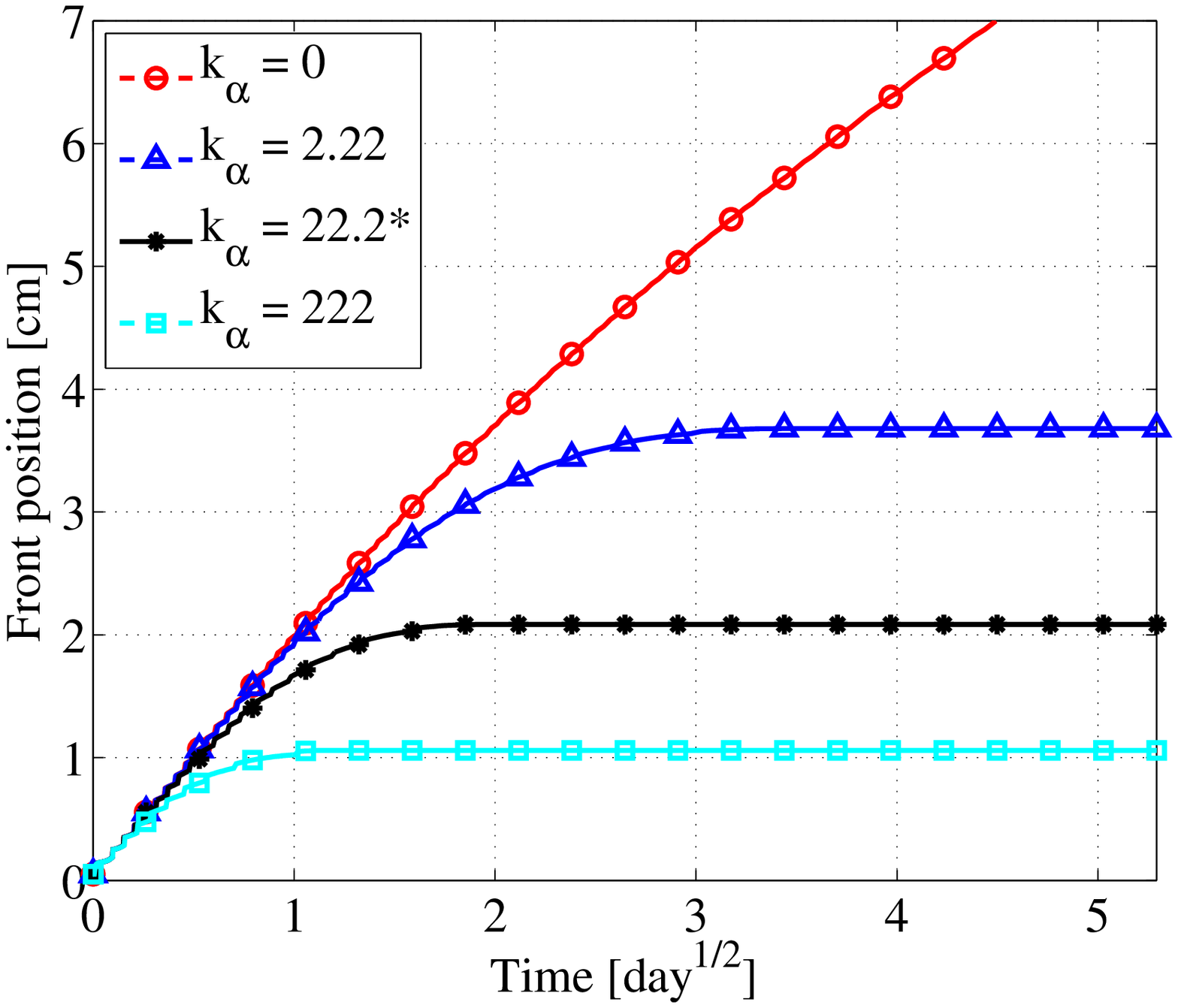}{0.45\textwidth} \\
    a.~~Final water content. & b.~~Wetting front position.
  \end{tabular}
  \caption{Water content and wetting front location for different values
    of the alite reaction rate, $k_\alite$.  In this and all succeeding
    figures, the base case is plotted using a solid black line and
    highlighted in the legend using ``*''.} 
  \label{fig:vary-ka}
\end{figure}

\begin{figure}[tbhp]
  \tabcapfont
  \begin{tabular}{c@{\hspace{0.1cm}}c}
    \myepsfile{runs/sumbetasat}{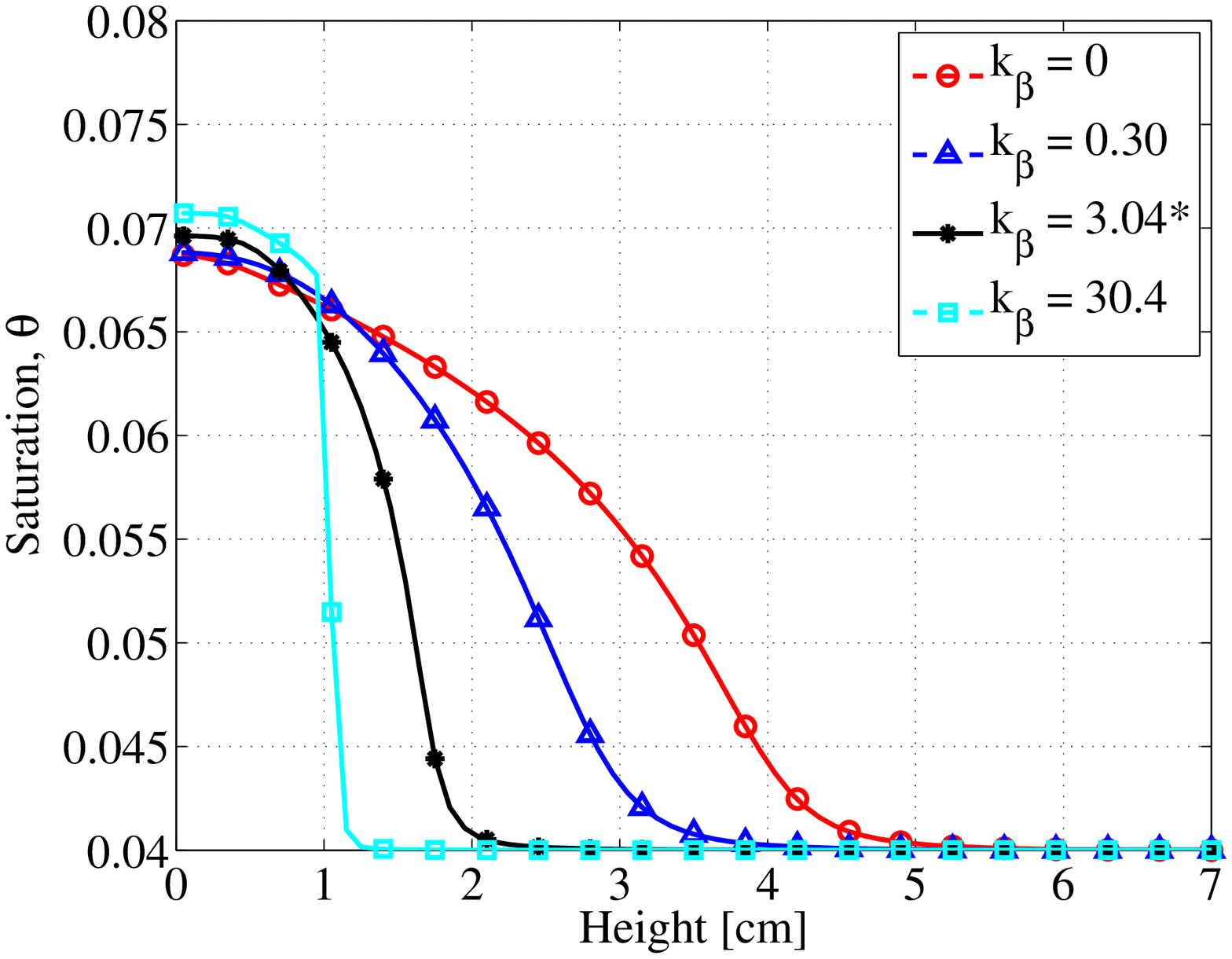}{0.50\textwidth} &
    \myepsfile{runs/sumbetapos}{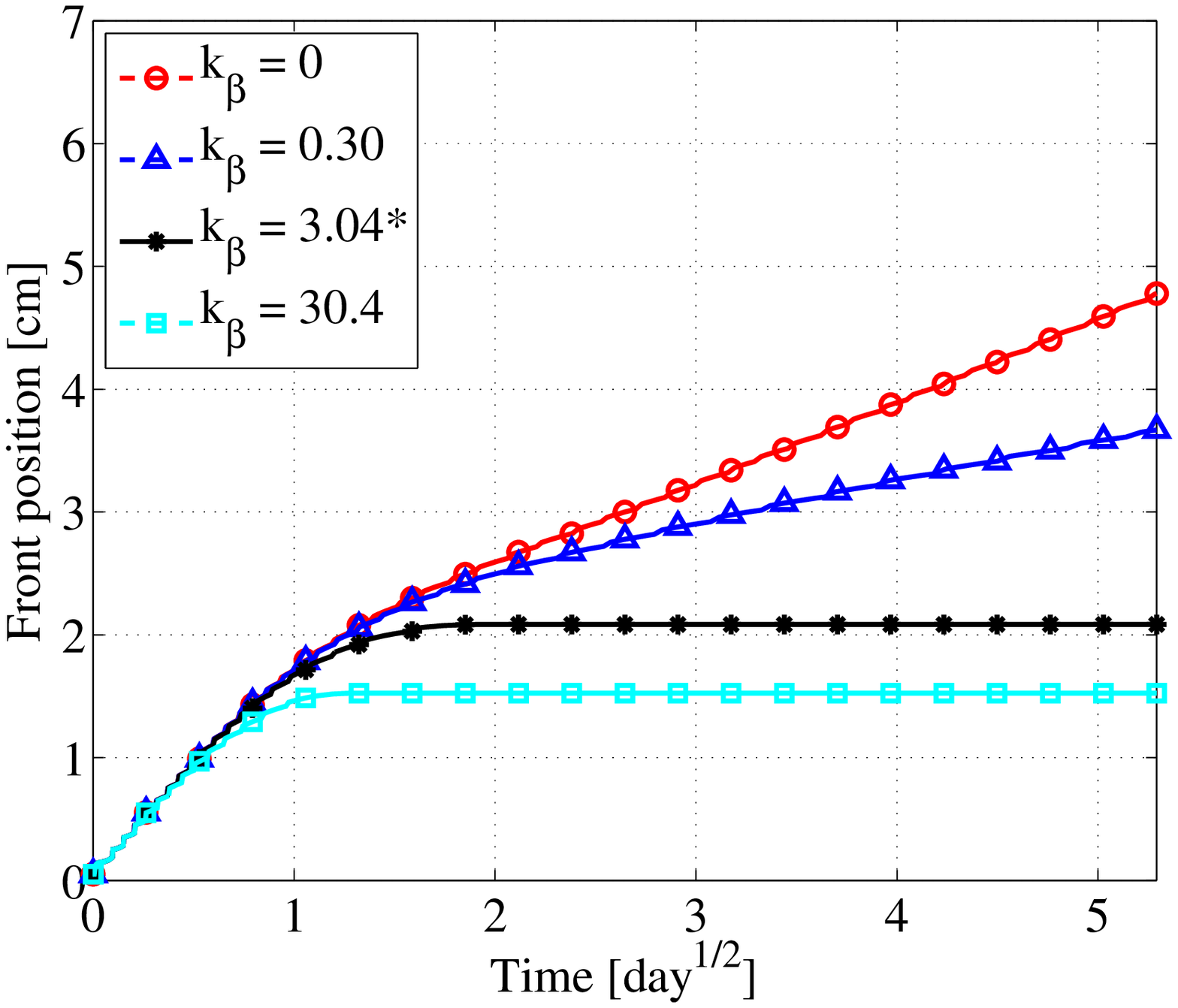}{0.45\textwidth} \\
    a.~~Final water content. & b.~~Wetting front position.
  \end{tabular}
  \caption{Water content and wetting front location for
    different values of the belite reaction rate, $k_\belite$.}
  \label{fig:vary-kb}
\end{figure}

In this section we vary the reaction rate parameters $k_\alite$ and
$k_\belite$ to investigate the effect of changes in the individual rates
as well as the relative importance of the two reaction routes leading to
production of \CSH gel.  To this end we hold $k_\belite$ constant and
scale $k_\alite$ by the factors 0, 0.1 and 10, and then repeat the same
procedure for $k_\belite$.  The resulting solutions are displayed in
Figs.~\ref{fig:vary-ka} and \ref{fig:vary-kb} from which we see that the
clogging seen in the final solution is very sensitive to changes in both
rates.  The results in both cases are similar, with effect of alite
being more pronounced; this is not surprising considering that the
initial concentration of alite is significantly larger than that of
belite (refer to values of $C_\alite^o$ and $C_\belite^o$ in
Table~\ref{tab:params}).  We also note that if the alite reaction rate
is taken small enough, then no stalling occurs and the wetting front
propagates essentially unhindered into the sample; the same is not true
of the belite rate since there is still enough alite being hydrated to
cause significant clogging.  The sensitivity to reaction rates
demonstrated by these results points to the importance of obtaining
accurate estimates of the rate parameters.


\subsection{Sensitivity to precipitation rate}
\label{sec:vary-kads}

\begin{figure}[tbhp]
  \tabcapfont
  \begin{tabular}{c@{\hspace{0.1cm}}c}
    \myepsfile{runs/sumadssat}{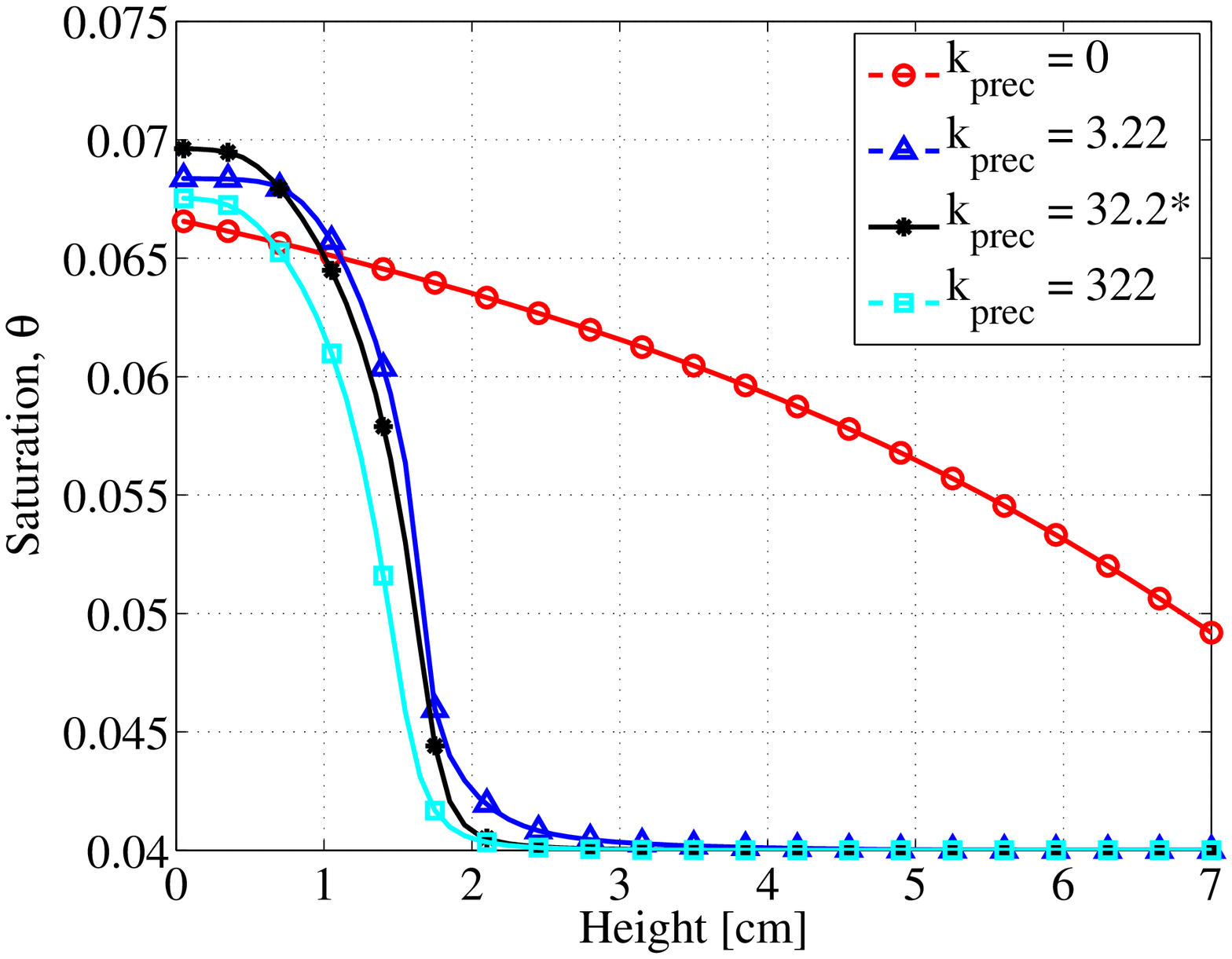}{0.50\textwidth} &
    \myepsfile{runs/sumadspos}{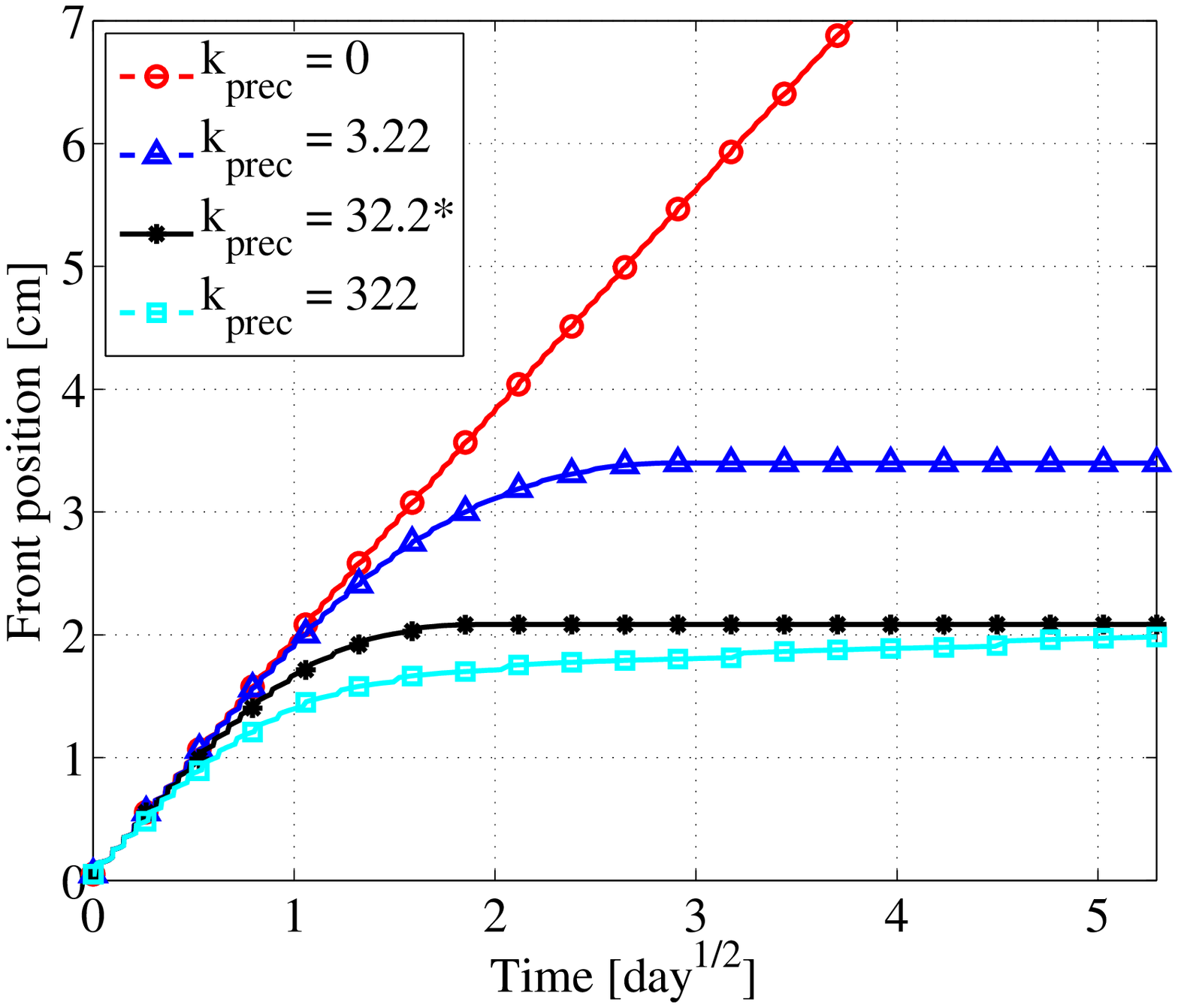}{0.45\textwidth} \\
    a.~~Final water content. & b.~~Wetting front position.
  \end{tabular}
  \caption{Water content and wetting front location for
    different values of the \CSH precipitation rate, $\kads$.}
  \label{fig:vary-kads}
\end{figure}

Since there is some uncertainty in the choice of the precipitation rate, it
is helpful to consider the effect of changes in $\kads$.  We ran three
additional simulations with $\kads=0.0, 3.22$ and $322$ and compared
those to the base case in Figure~\ref{fig:vary-kads}.  The $\kads=0$
case is identical to the case displayed in Fig~\ref{fig:base-sat}a (without
reactions) and from the remaining results it is clear that the solution
is relatively sensitive to the choice of precipitation rate.  We have done
our best to choose a value of $\kads$ consistent with \CSH precipitation
rates in the literature, but there is potentially much to be learned by
taking a more detailed look at the precipitation process
and including more details about this and other reaction mechanisms in
the model equations.

\subsection{Sensitivity to dissolution rate}
\label{sec:vary-kdes}

\begin{figure}[tbhp]
  \tabcapfont
  \begin{tabular}{c@{\hspace{0.1cm}}c}
    \myepsfile{runs/sumdessat}{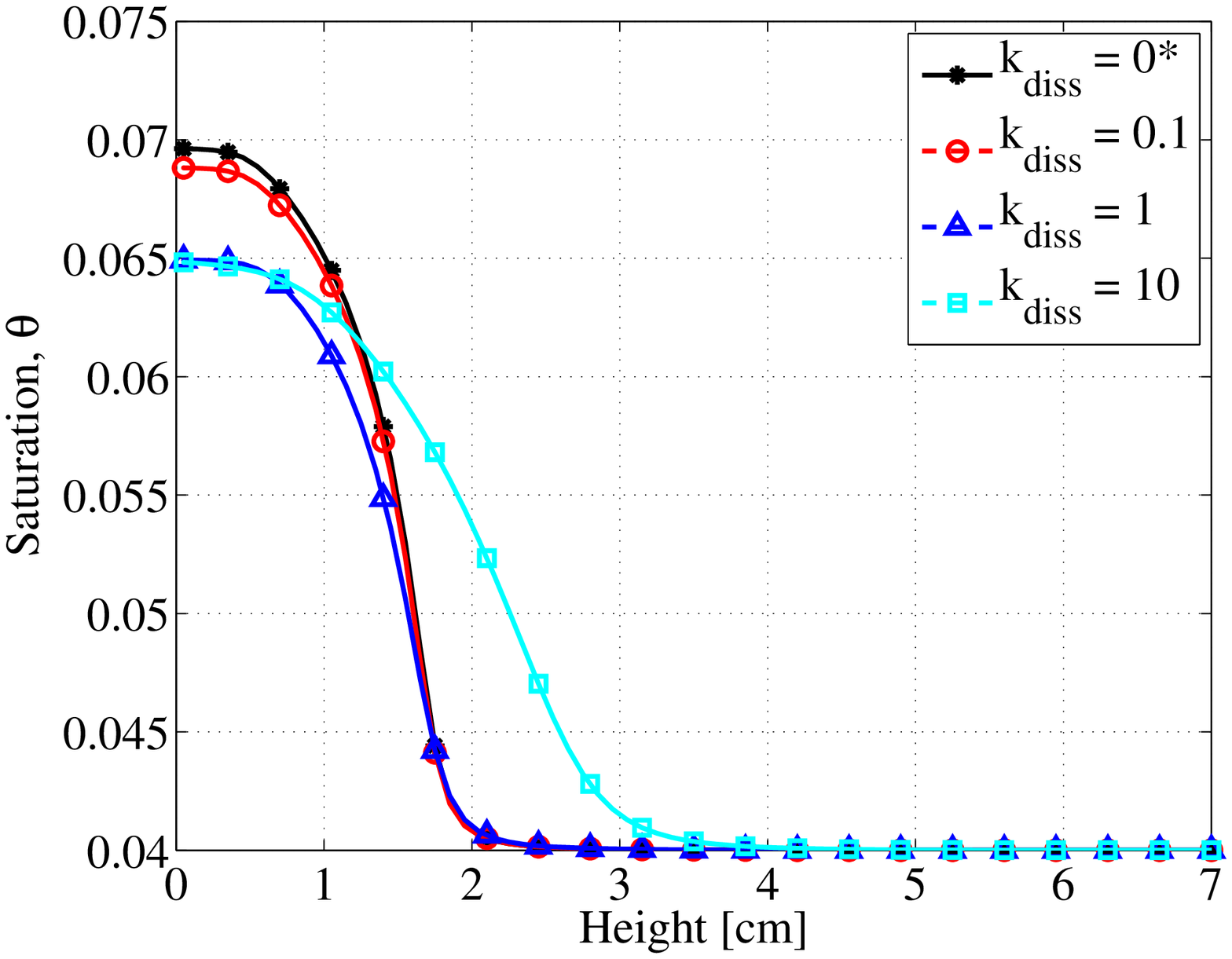}{0.50\textwidth} &
    \myepsfile{runs/sumdespos}{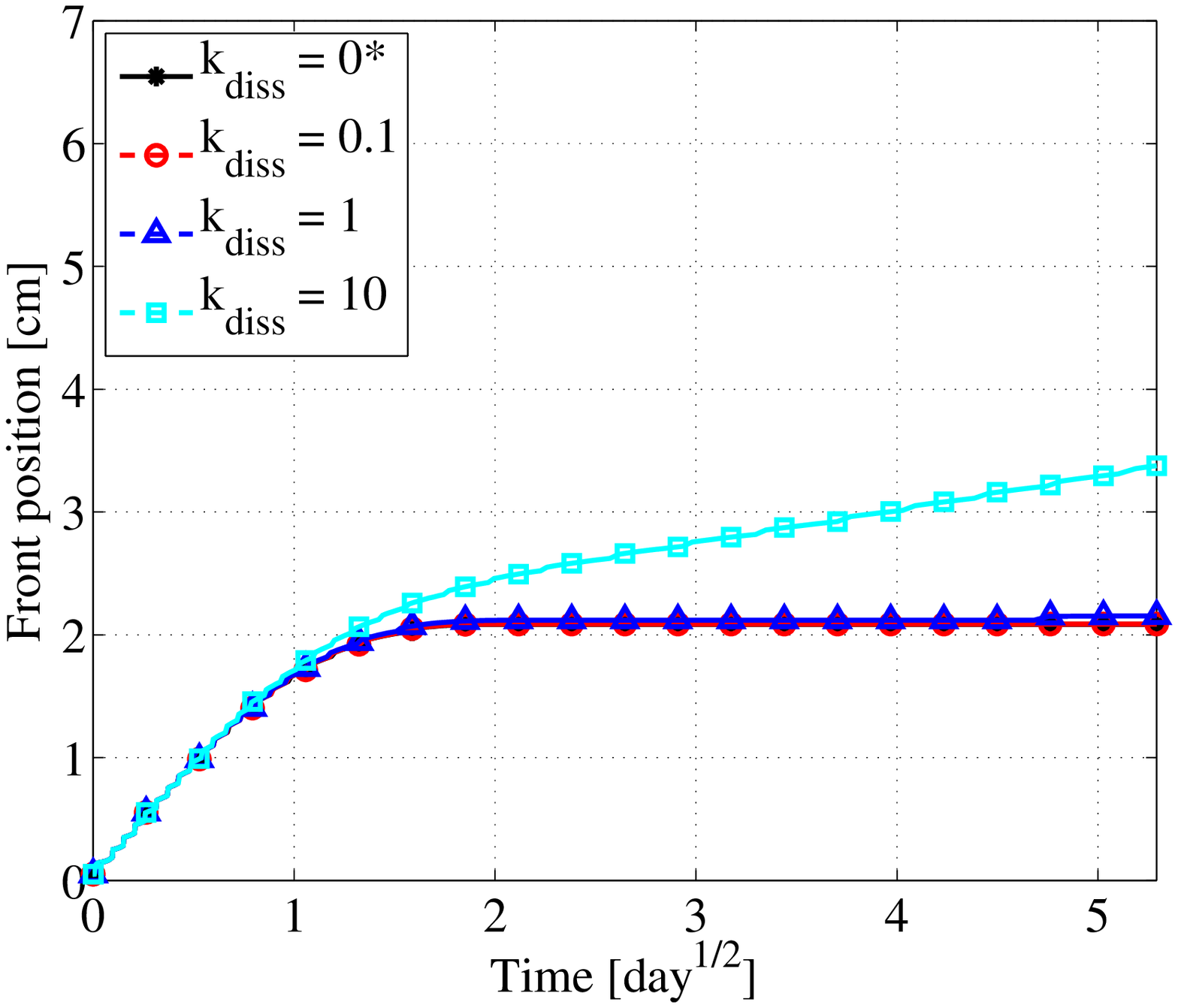}{0.45\textwidth} \\
    a.~~Final water content. & b.~~Wetting front position.
  \end{tabular}
  \caption{Water content and wetting front location for
    different values of the dissolution rate, $\kdes$.}
  \label{fig:vary-kdes}
\end{figure}

We have so far assumed that the formation of \CSH (gel) is an
irreversible process and no dissolution occurs, which is
consistent with assumptions made in many other models.  Since our focus
is on the phenomenon of re-wetting wherein time scales are much longer
than typically considered for initial hydration reactions, it is helpful
to consider the effect of incorporating a non-zero dissolution rate
constant $\kdes$.  To this end, we considered values of $\kdes=1$ and
and 10 $\mday^{-1}$ and compared the resulting solutions in
Fig.~\ref{fig:vary-kdes}, which clearly indicates that only for the
largest value of $\kdes$ is there any appreciable effect on the wetting
front position, although the water content does show some deviations at smaller
values of $\kdes$.  These results support our assumption that
dissolution has a negligible effect on the solution when
$\kdes\ll\kads$.

\subsection{Sensitivity to constituent diffusivity}
\label{sec:vary-Dabq}

\begin{figure}[tbhp]
  \tabcapfont
  \begin{tabular}{c@{\hspace{0.1cm}}c}
    \myepsfile{runs/sumdiffsat}{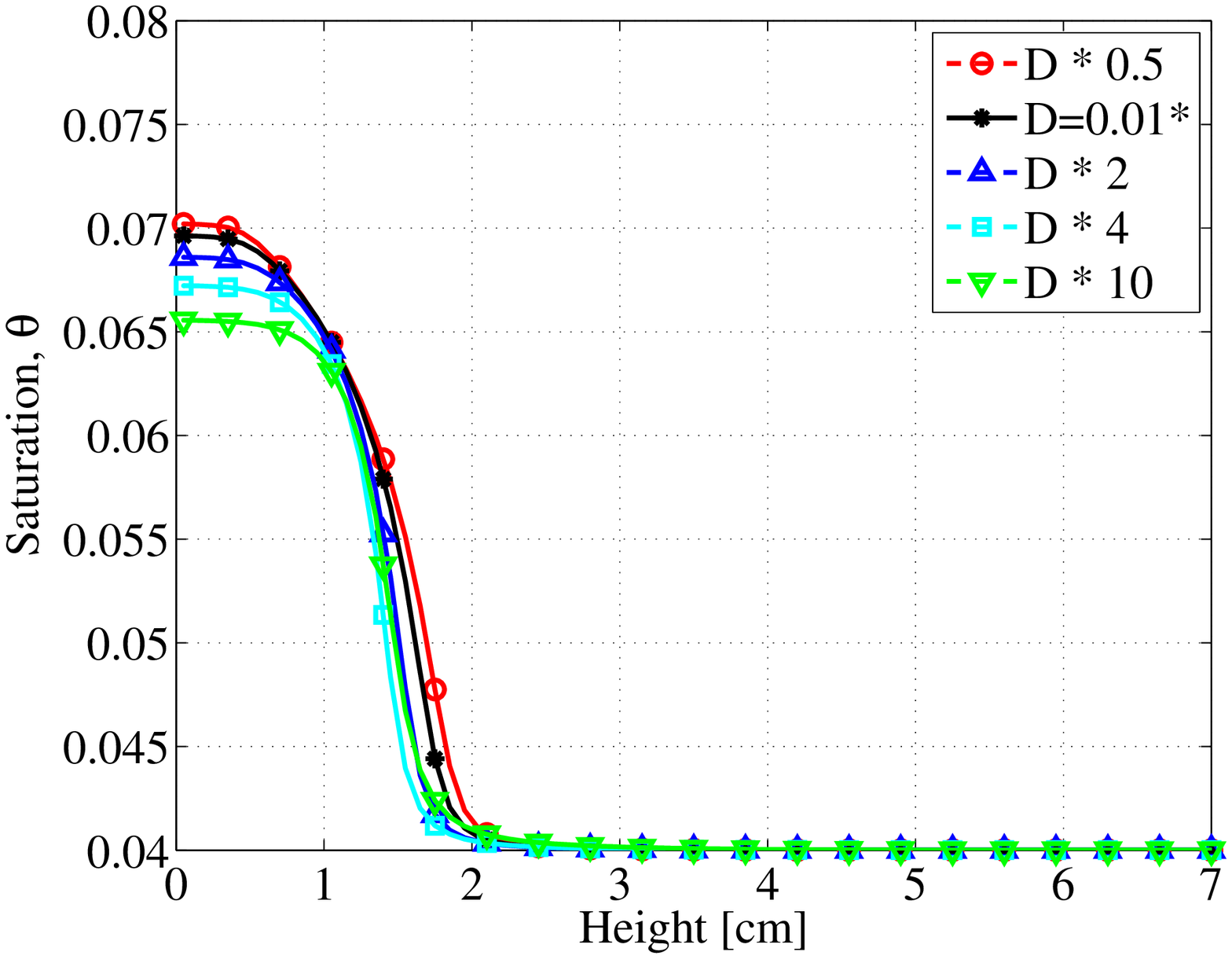}{0.50\textwidth} &
    \myepsfile{runs/sumdiffpos}{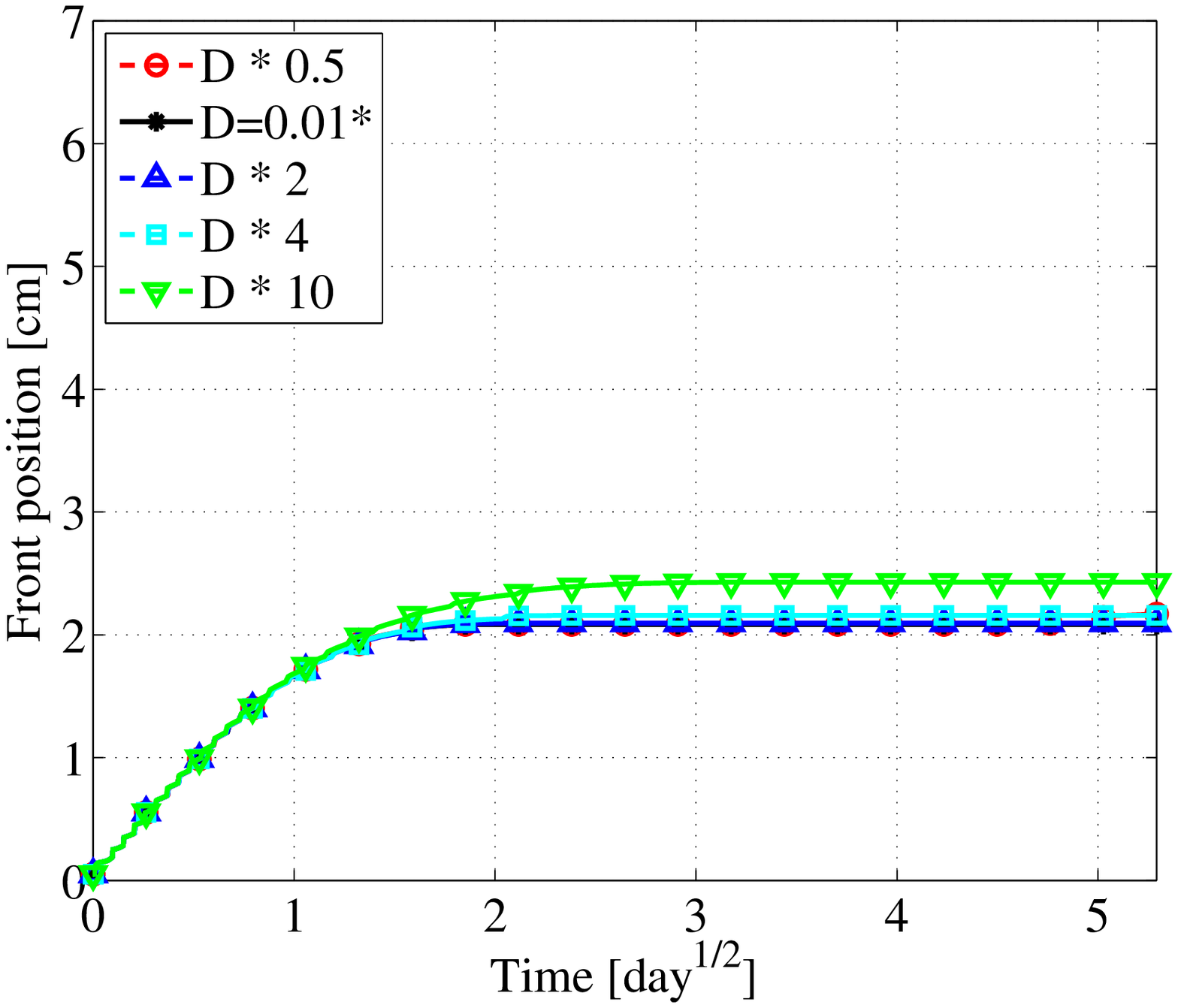}{0.45\textwidth} \\
    a.~~Final water content. & b.~~Wetting front position.
  \end{tabular}
  \caption{Water content and wetting front location
    obtained by varying the diffusivities $D_\alite$, $D_\belite$ and
    $D_\aq$.  In each case depicted, all three diffusivities are scaled 
    by the same constant factor.}
  \label{fig:vary-Dabq}
\end{figure}

We next investigate the effect of changing the diffusion coefficients
for the aqueous alite, belite and \CSH species.  We note that our model
ignores transport and reaction of individual ionic species and instead
approximates the diffusive transport by employing an \emph{effective
diffusion coefficient} for each constituent which may not be entirely
representative of how the individual ions would move in response to
concentration gradients in solution.  Fig.~\ref{fig:vary-Dabq}
demonstrates that changes in the diffusion coefficient by several orders
of magnitude have some effect on the steepness of the wetting front and
the distribution of constituents behind it, but have very little
influence on the location of the front itself. 

\subsection{Sensitivity to aggregate density}
\label{sec:vary-rhoagg}

\begin{figure}[tbhp]
  \tabcapfont
  \begin{tabular}{c@{\hspace{0.1cm}}c}
    \myepsfile{runs/sumaggsat}{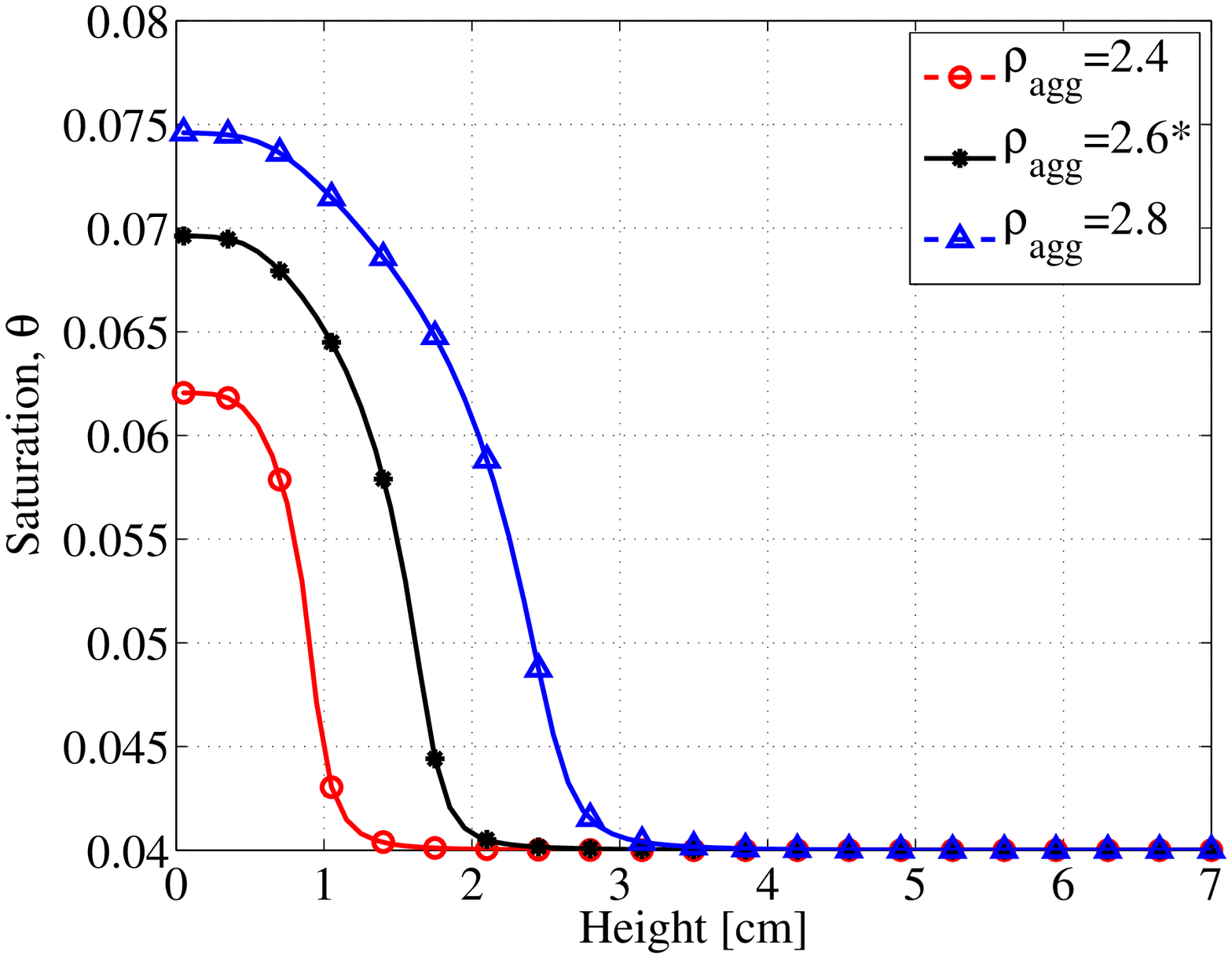}{0.50\textwidth} &
    \myepsfile{runs/sumaggpos}{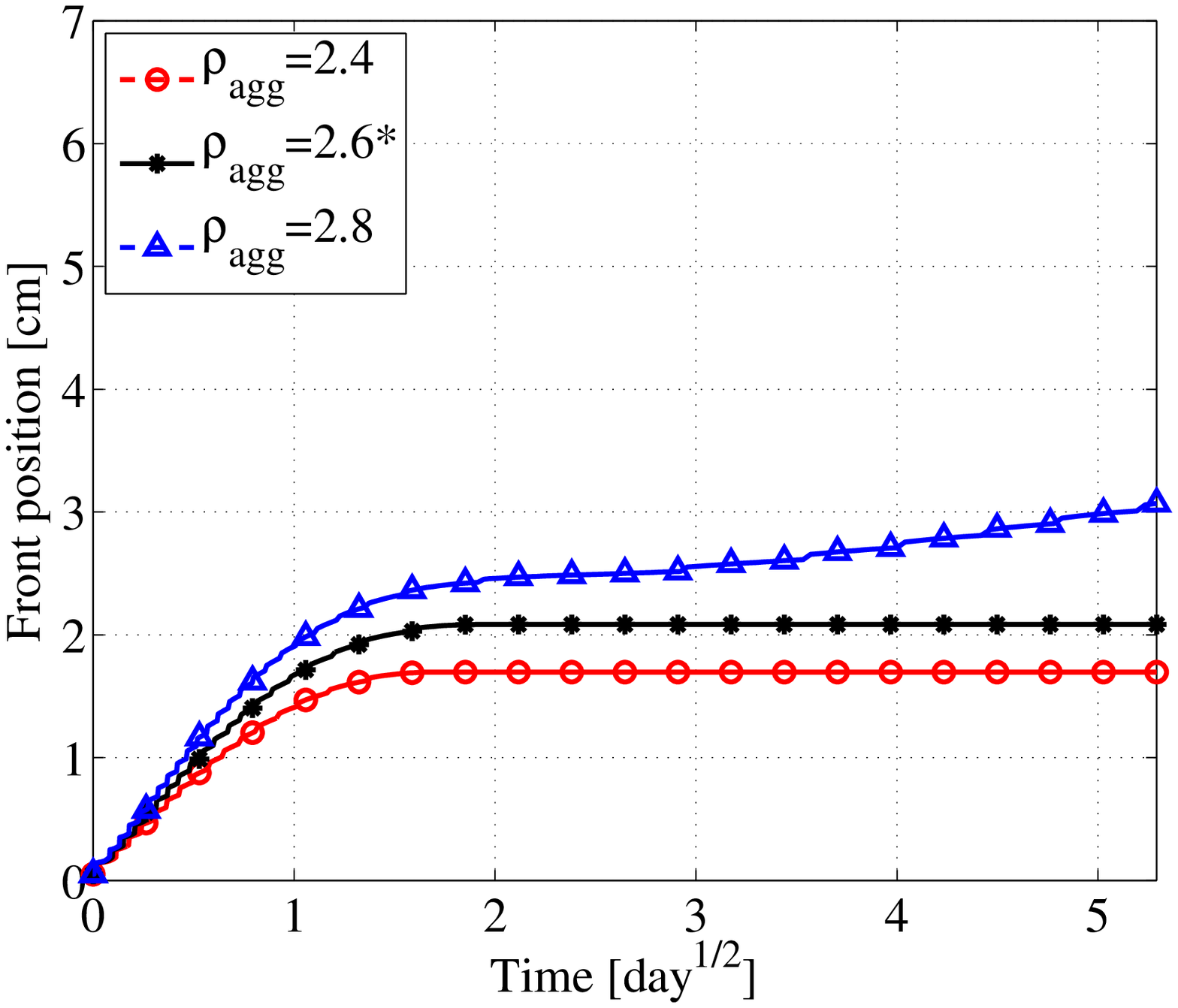}{0.45\textwidth} \\
    a.~~Final water content. & b.~~Wetting front position.
  \end{tabular}
  \caption{Water content and wetting front location for 
    different values of the aggregate density, $\rhoagg$.}
  \label{fig:rhoagg}
\end{figure}

The aggregate materials typically used in concrete include sand and
gravel of varying coarseness, all of which have
different density.  In practice, a combination of various aggregates is
frequently used and so we next investigate the effect of variations in
the aggregate density.  Fig.~\ref{fig:rhoagg} compares the solution when
$\rhoagg$ is varied between 2.4 and 2.8, and shows that even such
seemingly small changes in aggregate density can have a measurable
effect on clogging; in particular, as $\rhoagg$ increases, the degree of
clogging experienced decreases.  We therefore conclude that an
inaccurate value of the aggregate density parameter could lead to
incorrect results. 

\leavethisout{
\subsection{Sensitivity to \CSH gel density}
\label{sec:vary-rhogel}

\begin{figure}[tbhp]
  \tabcapfont
  \begin{tabular}{c@{\hspace{0.1cm}}c}
    \myepsfile{runs/sumgelsat}{sumgelsat.eps}{0.50\textwidth} &
    \myepsfile{runs/sumgelpos}{sumgelpos.eps}{0.45\textwidth} \\
    a.~~Final water content. & b.~~Wetting front position.
  \end{tabular}
  \caption{Water content and wetting front location for 
    different values of the \CSH gel density, $\rhoCSHgel$.}
  \label{fig:rhogel}
\end{figure}
}

\subsection{Effect of changes in cement mixture}
\label{sec:vary-cement}

Most concrete is mixed with a water-to-cement ratio $\Rwc$ lying
somewhere between 0.3 and 0.6.  It is well known that when $\Rwc$ is too
large the resulting concrete can be weak and so a smaller $\Rwc$ is
desirable in general.  On the other hand, if there is too little water
then the cement can become unworkable or there may even be insufficient
pore water to fully hydrate the silicates in the hydration process.
Consequently, optimizing concrete strength and durability requires a
fine tuning of the initial water content.  We have simulated the effect
of changes in composition by taking parameters as listed in
Table~\ref{tab:mixes}, which correspond to mixtures numbered 1 through 4
from \inlinecite{barrita-etal-2004}.  Outside of the variations in
$\Rwc$ and $\Rac$, a major difference between the various mixtures is
the presence of fly ash (in mixtures 2 and 3) or silica fume (in mixture
4).  Both of these low-density cement additives have the effect of
reducing the value of $\rhocem$, which in the case of mixtures 2 and 3
can change the resulting porosity $\porref$ significantly. 
\begin{table}[tbhp]
  \caption{Composition of cement mixtures taken
    from~{\protect\cite[Tab.~2]{barrita-etal-2004}}, with computed
    results compared in Fig.~{\protect\ref{fig:vary-mix}}.}
  \label{tab:mixes}
  \begin{tabular}{l@{\hspace{0.5cm}}|@{\hspace{0.5cm}}c@{\hspace{0.5cm}}c@{\hspace{0.5cm}}c@{\hspace{0.5cm}}c}\hline
    Mixture  & $\rhocem$ & $\Rwc$ & $\Rac$ & $\porref$ \\\hline
    1        & 3.15      & 0.599  & 5.39   & 0.113 \\
    2        & 2.62      & 0.364  & 3.13   & 0.074 \\  
    3 (base) & 2.83      & 0.333  & 2.86   & 0.066 \\  
    4        & 3.07      & 0.297  & 3.12   & 0.045 \\   
    \hline
  \end{tabular}
\end{table}

\begin{figure}[tbhp]
  \tabcapfont
  \begin{tabular}{c@{\hspace{0.1cm}}c}
    \myepsfile{runs/summixsat}{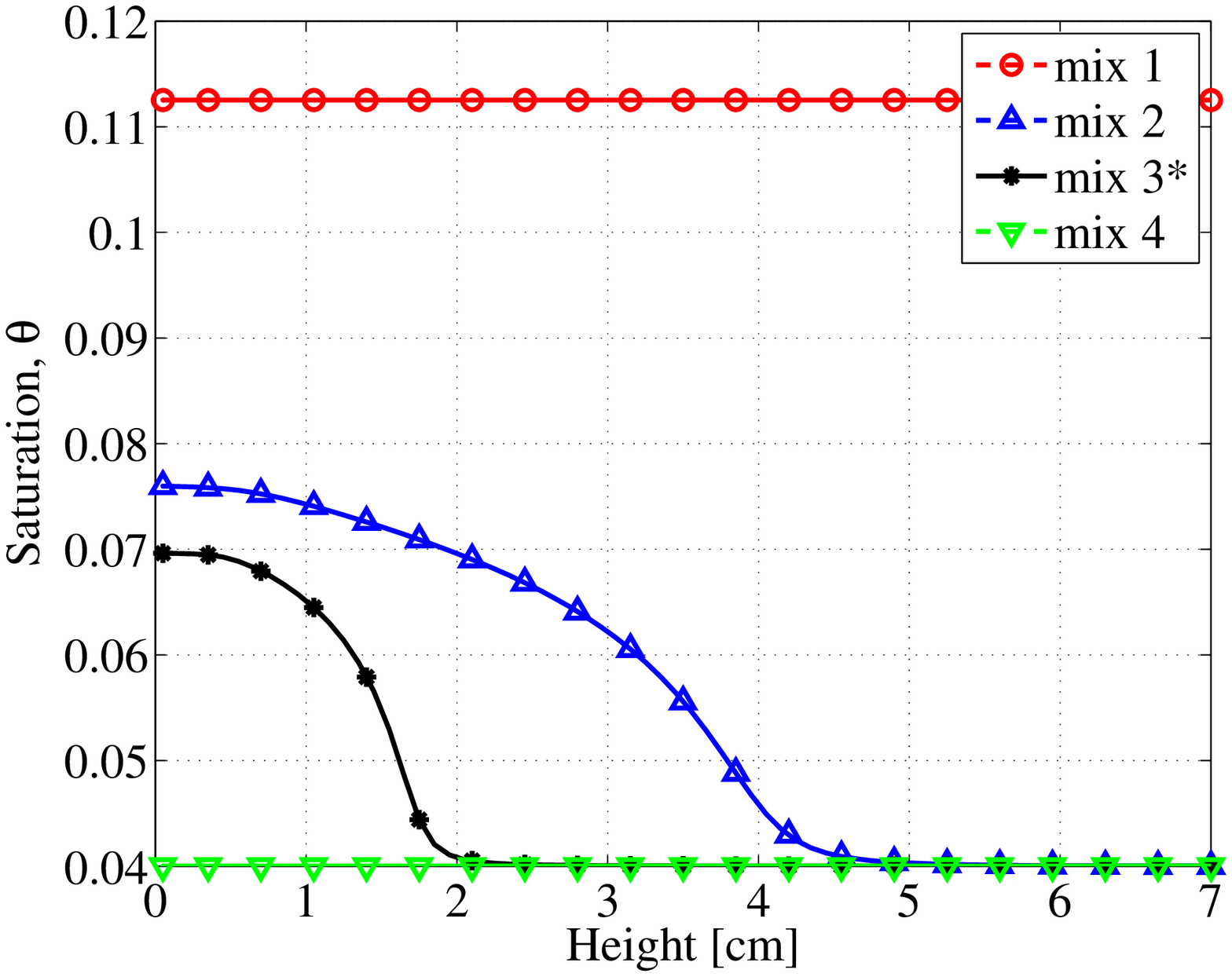}{0.50\textwidth} &
    \myepsfile{runs/summixpos}{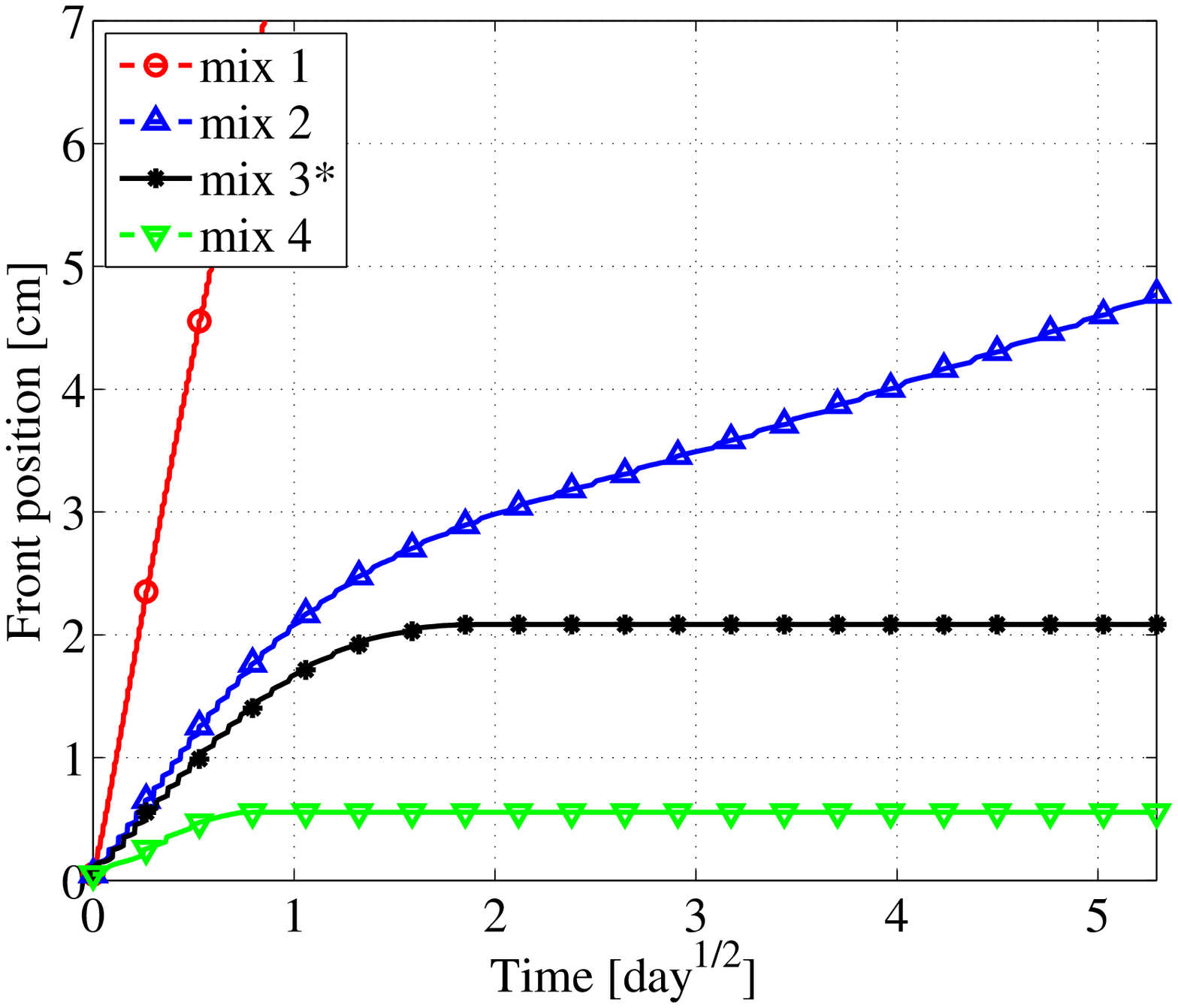}{0.45\textwidth} \\
    a.~~Final water content. & b.~~Wetting front position.
  \end{tabular}
  \caption{Water content and wetting front location obtained for various
    cement compositions, using mixtures 1--4
    in~{\protect\cite{barrita-etal-2004}}.}  
  \label{fig:vary-mix}
\end{figure}

The resulting numerical solutions are compared in
Fig.~\ref{fig:vary-mix} from which it is clear that the initial porosity
(as determined by the concrete mixture) can have a major impact on water
transport.  We note in particular that mixture 1 (with the largest value
of $\porref$) exhibits no clogging, while the low value of $\porref$ in
mixture 4 leads to very limited water transport, with the wetting front
stalling much closer to $x=0$.  Indeed, Barrita
\etal~\cite{barrita-etal-2004} observed in experiments that their
mixture 4 exhibited a much earlier onset of clogging than the other
concrete samples, an effect that is clearly captured in our simulations.
However, there remains some discrepancy in that experiments on mixture
1 exhibited a stalled wetting front, while our simulations show no
clogging in this case.

\leavethisout{
We performed one final comparison which attempts to take into account
the effect of curing time on the concrete.  Concrete which cures for a
longer period has more time to hydrate and therefore will tend to have a
higher degree of clogging than concrete cured for shorter periods.  The
fractional degree of hydration is well known to increase with the curing
time~\cite{tennis-jennings-2000} and so we have scaled the
base case values of $f_\alite=0.60$ and $f_\belite=0.20$ by factors 0.8
and 1.2 to see the effect of increasing during time.  The results are
plotted in Fig~\ref{fig:vary-fhyd}, from which we can infer that the
effect of changes in the fractional hydration on the resulting clogging
is actually quite weak.
\begin{figure}[tbhp]
  \tabcapfont
  \begin{tabular}{c@{\hspace{0.1cm}}c}
    \myepsfile{runs/sumhydsat}{sumhydsat.eps}{0.50\textwidth} &
    \myepsfile{runs/sumhydpos}{sumhydpos.eps}{0.45\textwidth} \\
    a.~~Final water content. & b.~~Wetting front position.
  \end{tabular}
  \caption{Water content and wetting front location obtained by
    varying the hydration fraction.}
  \label{fig:vary-fhyd}
\end{figure}
}

\section{Conclusions and future work}
\label{sec:conclusion}

We have developed a model for the transport and reaction of water
and other reactant species in hardened concrete subject to re-wetting.
Numerical simulations support our hypothesis that hydration of residual
silicates and subsequent formation of \CSH gel may be responsible for
the clogging phenomenon observed in experiments, which is the main 
contribution of this paper.

We investigated the sensitivity of the solution to changes in a
number of model parameters, from which we can conclude that the reaction
rate parameters (specifically $k_\alite$, $k_\belite$ and $\kads$) have
the most impact on the solution.  These are precisely the parameters
which are most difficult to ascertain owing to discrepancies in the
published literature, and in particular the lack of values for reaction
rates in actual concrete as opposed to idealized values obtained for
silicates prepared in aqueous solutions.   Consequently, more work is
required to ensure that inputs to our model are consistent with actual
concrete re-wetting scenarios. 

In addition to obtaining better estimates of the model parameters, there
are a number of extensions to the current model which may significantly
improve its predictive power.  We expect that the greatest impact may be
had by replacing the simple precipitation process embodied in our rate
parameter $\kads$ with a more realistic reaction mechanism that takes
into account details of the \CSH microstructure and hydration which
have recently been uncovered. Possible examples include:
\begin{itemize}
\item Incorporating the dynamics of individual ionic species through the
  addition of new transport equations and reaction kinetics along the
  lines of~\cite{preece-billingham-king-2001} or~\cite{meier-etal-2007}.
\item Investigating the hypothesis put forward
  in~\cite{tzschichholz-zanni-2001} that hydration kinetics is a
  two-stage process, consisting of an early accelerated hydration step
  followed by a slower hydration reaction that dominates in the longer
  term.  They suggest that this two-stage kinetics might arise from
  effects of either \CSH microstructure or precipitation kinetics,
  either of which could be considered in detail by appropriate
  modifications of our model.
\item Separating the \CSH gel into two forms characterized by different
  densities as suggested
  in~\cite{taylor-etal-1999,tennis-jennings-2000}, where the
  lower-density gel is thought to be primarily responsible for changes
  in porous structure.  Taylor \etal~\cite{taylor-etal-1999} also
  mention the importance of swelling in the cement matrix during initial
  cement hydration, which is an effect we have so far neglected.
\item Chemical shrinkage and the associated phenomenon of
  self-desiccation, which are known to have a significant impact on
  initial cement hydration~\cite{persson-1997}.
\end{itemize}

It may prove useful to incorporate other aspects of porous transport
that are commonly seen in modelling studies of ground water aquifers or
oil reservoirs, but have yet to be applied to the study of concrete.
For example, capillary hysteresis has been identified as an important
aspect of cement hydration~\cite{beaudoin-1999} and results from the
soil sciences community~\cite{hillel-1998,kool-parker-1987} could
certainly be applied in this context.  The issues raised
in~\cite{gray-miller-2004} surrounding the impact of variable porosity
on models of multi-phase transport should also be applicable to cement
and concrete.  Our model can be easily adapted to study other stages in
the life of concrete such as initial hydration, carbonation, aging or
degradation.  Finally, the approach we have developed here would also be
applicable to the study of other transport phenomena such as polymer
flooding in enhanced oil recovery, where chemical reactions and
solution-dependent parameters are important.

\leavethisout{
  Key numerical issues to sort out here:
  \begin{itemize}
  \item {\LARGE $\ast\ast\ast$} Look at the effect of using an analytic
    Jacobian vs. a finite difference approximation.  Or alternately we
    could use the ``poor man's automatic differentiation'' or PMAD
    approach advocated by Shampine~\cite{shampine-2007}.
  \item Extend the analysis or the numerical method
    from~\cite{mitchell-morton-spence-2006} to deal with the reaction
    terms in our concrete transport model.  We might be able to use this
    to derive a method which is guaranteed to be mass-conserving and/or
    positive.
  \item Apply the methods developed in
    \cite{divisek-fuhrmann-gartner-jung-2003} or
    \cite{shampine-sommeijer-verwer-2006} to this and other more
    complicated concrete chemistry problems.
  \end{itemize}

  Other ideas:
  \begin{itemize}
  \item Revisit once more the term $\satfac$ in the reaction terms and
    check whether it shouldn't really be only $\sat$ (see Meier+, 2005).
  \end{itemize}
}

\ifthenelse{\boolean{@PlainVersion}}{%
  \section*{Acknowledgements}}{%
  \begin{acknowledgements}}
  
  This work was supported by grants from the Natural Sciences and
  Engineering Research Council of Canada and the MITACS Network of
  Centres of Excellence.  JMS was supported by a Research Fellowship
  from the Alexander von Humboldt Foundation during a visit to the
  Fraunhofer Institut Techno- und Wirtschaftsmathematik in
  Kaiserslautern.  We thank Dr.\ Jes\'us Cano Barrita (Instituto
  Polit\'ecnico Nacional, Oaxaca, Mexico) for many insightful
  discussions and for providing the experimental data.  We are also
  sincerely grateful to the four anonymous referees whose extensive
  comments have significantly improved this work.
  
  \ifthenelse{\boolean{@PlainVersion}}{}{\end{acknowledgements}}

\ifthenelse{\boolean{@EDMGRVersion}}{
  

}{%
  \ifthenelse{\boolean{@PlainVersion}}{
    \bibliographystyle{plain}%
  }{%
    \bibliographystyle{klunumunsrt}}
  \bibliography{short-abbrevs,concrete,soils,ballard,numanal,hyperbolic,mybooks}
}

\end{document}